\documentclass[12pt,preprint]{emulateapj}
\usepackage{amsmath}
\usepackage{soul}
\begin{document}

\title{$z\gtrsim7$ Galaxies with Red Spitzer/IRAC [3.6]$-$[4.5] colors
  in the full CANDELS data set: The brightest-known galaxies at
  $z\sim7$-9 and a probable spectroscopic confirmation at $z=7.48$}

\author{G. W. Roberts-Borsani\altaffilmark{1,2}, R. J. Bouwens\altaffilmark{1},
        P. A. Oesch\altaffilmark{3}, I. Labbe\altaffilmark{1}, R. Smit\altaffilmark{1,4}, G. D. Illingworth\altaffilmark{5}, P. van Dokkum\altaffilmark{3}, B. Holden\altaffilmark{5}, V. Gonzalez\altaffilmark{6}, M. Stefanon\altaffilmark{1}, B. Holwerda\altaffilmark{1}, S. Wilkins\altaffilmark{7}}

\altaffiltext{1}{Leiden Observatory, Leiden University, NL-2300 RA Leiden, Netherlands}
\altaffiltext{2}{Department of Physics and Astronomy, University College London, Gower Street, London WC1E 6BT, UK}
\altaffiltext{3}{Department of Astronomy, Yale University, New Haven, CT 06520}
\altaffiltext{4}{Department of Physics and Astronomy, South Road, Durham, DH1 3EE, United Kingdom}
\altaffiltext{5}{UCO/Lick Observatory, University of California, Santa Cruz, CA 95064}
\altaffiltext{6}{University of California, Riverside, CA 92521, USA}
\altaffiltext{7}{Department of Physics \& Astronomy, University of Sussex, Falmer, BRIGHTON, BN1 9QH, United Kingdom}

\begin{abstract}
We identify 4 unusually bright (H$_{160,AB}$\textless25.5) galaxies
from HST and Spitzer CANDELS data with probable redshifts
\textit{z}$\sim$7-9. These identifications include the
brightest-known galaxies to date at $z\gtrsim7.5$.  As $Y$-band
observations are not available over the full CANDELS program to
perform a standard Lyman-break selection of $z>7$ galaxies, we
employ an alternate strategy using deep Spitzer/IRAC data. We
identify z$\sim$7.1-9.1 galaxies by selecting \textit{z}$\gtrsim$6
galaxies from the HST CANDELS data that show quite red IRAC
[3.6]$-$[4.5] colors, indicating strong [OIII]+H$\beta$ lines in the
4.5$\mu$m band.  This selection strategy was validated using a modest
sample for which we have deep Y-band coverage, and subsequently used to 
select the brightest $z\geq7$ sources.  Applying the
IRAC criteria to all HST-selected optical-dropout galaxies over the
full $\sim$900 arcmin$^{2}$ of the CANDELS survey revealed four
unusually bright $z\sim7.1$, 7.6, 7.9 and 8.6 candidates. The median
[3.6]$-$[4.5] color of our selected $z\sim7.1$-9.1 sample is
consistent with rest-frame [OIII]+H$\beta$ EWs of $\sim$1500\AA$\,$ in
the [4.5] band.  Keck/MOSFIRE spectroscopy has been independently
reported for two of our selected sources, showing Ly$\alpha$ at 
redshifts of 7.7302$\pm$0.0006 and 8.683$_{-0.004}^{+0.001}$, respectively. 
We present similar Keck/MOSFIRE spectroscopy for a third selected galaxy 
with a probable 4.7$\sigma$ Ly$\alpha$ line at $z_{spec}=$7.4770$\pm$0.0008. 
All three have H$_{160}$-band magnitudes of $\sim$25 mag and are $\sim$0.5 
mag more luminous ($M_{1600}\sim-22.0$) than any previously discovered 
\textit{z}$\sim$8 galaxy, with important implications for the UV LF. Our 3 
brightest, highest redshift $z$\textgreater7 galaxies all lie within the 
CANDELS EGS field, providing a dramatic illustration of the potential 
impact of field-to-field variance.
\end{abstract}


\section{Introduction}
\label{sec:intro}

The first galaxies are believed to have formed within the first
300-400 Myr of the Universe and great strides have been made towards
identifying objects within this era.  Since the installation of the
Wide Field Camera 3 (WFC3) instrument on the Hubble Space Telescope
(HST), an increasing number of candidates have been identified by
means of their photometric properties, with $\gtrsim$700 probable
galaxies identified at $z\sim$7-8 (\citealt{bouwens15}: see also
\citealt{mclure2013}; \citealt{schenker2013}; \citealt{lorenzoni2013};
\citealt{schmidt2014}; \citealt{bradley14}; \citealt{mason2015};
\citealt{finkelstein15}; \citealt{atek15}) and another 10-15
candidates identified even further out at $z\sim$9-11 (e.g.,
\citealt{zheng12}; \citealt{ellis13}; \citealt{oesch14a,oesch14b};
\citealt{bouwens15}; \citealt{zitrin14}; \citealt{zheng14};
\citealt{ishigaki15}; \citealt{mcleod14}).

One of the most interesting questions to investigate with these large
samples is the build-up and evolution of galaxies.  While these issues
have long been explored in the context of fainter galaxies through the
evolution of the $UV$ LF, less progress has been made in the study of
the most luminous galaxies due to the large volumes that must be
probed to effectively quantify their evolution.

The entire enterprise of finding especially bright galaxies at
$z\geq7$ has been limited by the availability of sufficiently deep,
multi-wavelength near-infrared data over wide areas of the sky.  The
most noteworthy such data sets are the UKIDSS UDS program (Lawrence et
al.\ 2007), the UltraVISTA program (McCracken et al.\ 2012), the
902-orbit CANDELS program from the Hubble Space Telescope (Grogin et
al.\ 2011; Koekemoer et al.\ 2011), the BoRG/HIPPIES pure-parallel
data set (Trenti et al.\ 2011; Yan et al.\ 2011; Bradley et al.\ 2012;
Schmidt et al.\ 2014; Trenti 2014), and the ZFOURGE data set (Tilvi et
al.\ 2013; I. Labb{\'e} et al.\ 2016, in prep)

Of these surveys, arguably the program with the best prospects for
probing the bright end of the $z>7$ population would be the wide-area
CANDELS program.\footnote{In principle, the wide-area ($\sim$1
  deg$^2$) UDS and UltraVISTA programs have great potential to find
  large numbers of bright $z\gtrsim6$ sources as demonstrated by the
  recent Bowler et al.\ (2014) results (see also Bowler et al.\ 2015),
  but may not yet probe deep enough to sample the $z\gtrsim8$ galaxy
  population.}  The challenge with CANDELS has been that it is only
covered with particularly deep near-infrared observations from
1.2$\mu$m to 1.6$\mu$m, but lacks HST-depth $Y$-band observations at
1.05$\mu$m over the majority of the area.  Deep observations at
1.05$\mu$m are needed for the determination of photometric redshifts
for galaxies in the redshift range $z\sim6.3$ to $z\sim8.5$.  While
this can be partially compensated for by the availability of
moderate-depth $1.05\mu$m observations from various ground-based
programs over the CANDELS program, e.g., HUGS \citep{fontana14},
UltraVISTA \citep{mccracken12}, and ZFOURGE (I. Labb{\'e} et al. 2016,
\textit{in prep}), such observations are not available over the entire
program, making it difficult to consider a search for bright $z>7$
galaxies over the full area.

Fortunately, there appears to be one attractive, alternate means for
making use of the full CANDELS area to search for bright $z>7$
galaxies.  This is to exploit the availability of uniformly deep
Spitzer/IRAC observations over the full area (e.g., Ashby et
al.\ 2013) and redshift information present in the [3.6]$-$[4.5]
colors of $z\sim5$-8 galaxies.  As demonstrated by many authors (e.g.,
\citealt{labbe13}; \citealt{smit14a,smit14b}; \citealt{bowler14};
\citealt{laporte14,laporte15}; \citealt{huang16}), the [3.6]$-$[4.5]
colors appear to depend on redshift in a particularly well-defined
way, a dependence which appears to arise from very strong nebular
emission lines, such as H$\alpha$ and [OIII]$\lambda$5007 \AA, which
pass through the IRAC bands at particular redshifts.  For example,
while $z\sim$6.8 galaxies have very blue [3.6]$-$[4.5] colors likely
due to contamination of the [3.6] filter by [OIII]+H$\beta$ lines (and
no similar contamination of the [4.5] band), $z\geq$7 galaxies exhibit
much redder [3.6]$-$[4.5] colors, as only the 4.5$\mu$m band is
contaminated by the especially strong [OIII]+H$\beta$ lines
(\citealt{labbe13}; \citealt{wilkins13}; \citealt{smit14a}).

Here, we make use of the redshift information in the Spitzer/IRAC
observations and apply a consistent set of selection criteria to
search for bright $z\sim8$ galaxies over all 5 CANDELS fields.  A full
analysis of the HST + ground-based observations is made in
preselecting candidate $z\gtrsim6$ galaxies, for further consideration
with the available Spitzer/IRAC data.  The identification of such
bright sources allows us to better map out the bright end of the UV
luminosity function (LF) at $z>7$ and constrain quantities like the
characteristic luminosity $M^*$ or the functional form of the LF at
$z>7$.  \citet{bouwens15} only observe a modest ($\sim$0.6$\pm$0.3
mag) brightening in the characteristic luminosity $M^*$ -- or bright
end cut-off -- from $z\sim8$ to $z\sim5$ taking advantage of the full
CANDELS + XDF + HUDF09-Ps search area ($\sim$1000 arcmin$^2$).
\citet{bowler15} also report evidence for a limited evolution in the
characteristic luminosity with cosmic time, based on a wider-area
search for $z\sim6$-7 galaxies found over the $\sim$1.7 deg$^2$
UltraVISTA+UDS area.  Limited evolution was also reported by
Finkelstein et al.\ (2015) in subsequent work, but utilizing a
$\sim$3-15$\times$ smaller area than Bouwens et al.\ (2015) or Bowler
et al.\ (2015) had used.

This paper is organised as follows:  \S2 presents our $z\sim$5-8
catalogs and data sets, as well as methodology for performing
photometry. \S3 describes the selection criteria we define for our
samples and methodology.  \S4 presents the results of our
investigation and discusses the constraints added by $Y$-band
observations and Keck/MOSFIRE spectroscopy.  In \S5, we use the
present search results to set a constraint on the bright end of the
$z>7$ LF.  Finally, \S6 includes a summary of our paper and a
prospective.  Throughout this paper, we refer to the HST F606W, F814W,
F105W, F125W, F140W, and F160W bands as \textit{V}$_{606}$,
\textit{I}$_{814}$, \textit{Y}$_{105}$, \textit{J}$_{125}$,
\textit{JH}$_{140}$ and \textit{H}$_{160}$, respectively, for
simplicity. We also assume \textit{H}$_{0}$ = 70 km/s/Mpc, $\Omega_{m}
=$ 0.3, and $\Omega_{\wedge} =$ 0.7. All magnitudes are in the AB
system \citep{oke83}.

\begin{deluxetable}{cccccc}
\tablewidth{0pt}
\tablecolumns{6}
\tabletypesize{\footnotesize}
\tablecaption{Summary of Data Sets Utilized in Current Search.\label{tab:datasets}}
\tablehead{
\colhead{} & \colhead{} & \multicolumn{4}{c}{Depth ($5\sigma$)} \\
\colhead{Data Set} & 
\colhead{Area} & 
\colhead{$J_{125}$} &
\colhead{$H_{160}$} &
\colhead{$[3.6]$} &
\colhead{$[4.5]$}}
\startdata
CANDELS GS DEEP & 64.5 & 27.8 & 27.5 & 26.1 & 25.9 \\
CANDELS GS WIDE & 34.2 & 27.1 & 26.8 & 26.1 & 25.9 \\
ERS & 40.5 & 27.6 & 27.4 & 26.1 & 25.9 \\
GS other\tablenotemark{a} & 31.8 \\
CANDELS GN DEEP & 62.9 & 27.7 & 27.5 & 26.1 & 25.9 \\
CANDELS GN WIDE & 60.9 & 26.8 & 26.7 & 26.1 & 25.9 \\
GN other\tablenotemark{a} & 34.0 \\
CANDELS UDS     & 191.2\tablenotemark{b} & 26.6 & 26.8 & 25.5 & 25.3 \\
CANDELS COSMOS  & 183.9\tablenotemark{b} & 26.6 & 26.8 & 25.4 & 25.2 \\
CANDELS EGS     & 192.4\tablenotemark{b} & 26.6 & 26.9 & 25.5 & 25.3 \\
Total &       896.3 
\enddata
\tablenotetext{a}{Photometry over a 31.8 and 34 arcmin$^2$ area within
  the GS and GN fields is not available in the Bouwens et al.\ (2015)
  catalogs, due to these catalogs only including regions which have
  $\gtrsim70$\% of the full depth available in the $B_{435}$,
  $V_{606}$, $i_{775}$, $z_{850}$, $Y_{105}$, $J_{125}$, and $H_{160}$
  bands.  For these regions, we make use of the Skelton et
  al.\ (2014) catalogs to search for $z\geq7$ galaxies.}
\tablenotetext{b}{The Bouwens et al.\ (2015) catalogs only cover the
  $\sim$450 arcmin$^2$ region from the CANDELS-UDS, COSMOS, and EGS
  fields where deep ACS and WFC3/IR data are available from CANDELS
  (75\% of the area).  In searching the CANDELS UDS, COSMOS, and EGS
  fields for $z\sim8$ candidates, we make use of the Bouwens et
  al.\ (2015) catalogs where available and the Skelton et
  al.\ (2014) catalogs otherwise.}
\end{deluxetable}

\begin{figure}
\epsscale{1.25}
\plotone{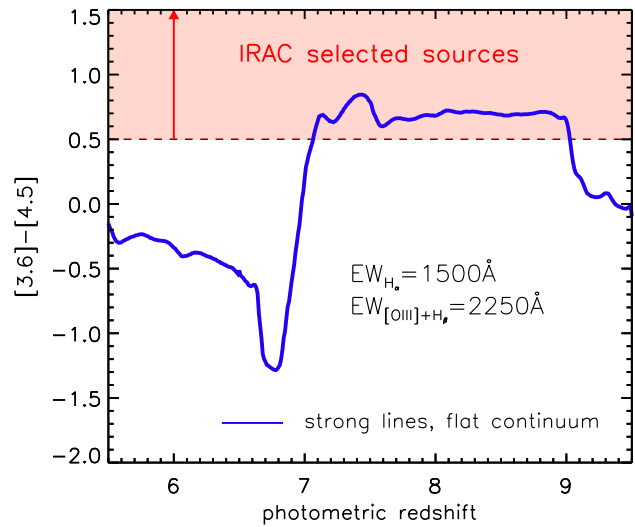}
\caption{Spitzer/IRAC [3.6]$-$[4.5] color vs. photometric redshift plot
  for young ($\sim$5 Myr) stellar population with very strong nebular
  emission lines (EW$_{H\alpha}$=1500$\AA$) and a flat continuum.
  Also assumed are fixed flux ratios between emission lines from Table
  1 of \citet{anders03} for 0.2 Z$_\odot$ metallicity, whilst assuming
  case B recombination for the H$\alpha$/H$\beta$ flux ratio.  The
  $[3.6]-[4.5]$ color of galaxies is expected to become quite red at
  $z\gtrsim7$ due to the impact of the [OIII] line on the 4.5$\mu$m
  band and no comparably bright nebular line in the
  3.6$\mu$m.\label{fig:color}}
\end{figure}

\section{Observational data sets, photometry and z$\sim$5-8 sample}

\subsection{HST + ground-based data set and photometry}

The sample of $z\sim8$ galaxies we identify in this paper is based on
HST + ground-based observations that were acquired over 5 CANDELS and
ERS fields (\citealt{grogin11}; \citealt{koe11}; Windhorst et
al.\ 2011).

The near-IR HST observations over the CANDELS fields range in depth
from $\sim$4 orbits over the $\sim$130 arcmin$^2$ CANDELS DEEP
components in GOODS-North (GN) and GOODS-South (GS) to $\sim1$ orbit
depth over the $\sim$550 arcmin$^2$ CANDELS WIDE component in the GN,
GS, UDS, COSMOS, and EGS fields.  Over the GN and GS fields, the
near-IR imaging observations are available in the $Y_{105}$,
$J_{125}$, and $H_{160}$ bands, while in the UDS, COSMOS, and EGS
fields, the near-IR observations are available in the $J_{125}$ and
$H_{160}$ bands.

These fields also feature observations at optical wavelengths with the
HST ACS camera in the $B_{435}$, $V_{606}$, $i_{775}$, $I_{814}$ and
$z_{850}$ bands for CANDELS-GN+GS (with 3-10+ orbits per band), as
well as $V_{606}$ and $I_{814}$ observations ($\sim$2-orbit depth) for
the CANDELS-UDS+COSMOS+EGS fields. 

In addition to the HST observations, these fields also have very deep
ground-based observations from CFHT, Subaru Suprime-Cam, VLT HAWK-I,
and VISTA/VIRCAM over the latter fields.  Optical data are available
in CANDELS-COSMOS field in the \textit{u}, \textit{g}, \textit{r},
\textit{i}, \textit{y} and \textit{z} bands as part of the CFHT legacy
survey, and also in the \textit{B}, \textit{g}, \textit{V},
\textit{r}, \textit{i} and \textit{z} bands from Subaru observations
over the same field (\citealt{capak11}).  The CANDELS-EGS field is
observed in the same bands as the COSMOS field, as part of the CFHT
legacy survey, whilst the CANDELS-UDS field is observed by Subaru as
part of the Subary XMM-Newton Deep Field (SXDF) program (Furusawa et
al.\ 2008).  For extended sources, these optical observations reach
similar or greater depths to the available HST data over these fields
(i.e., 26 mag to 28 mag at $5\sigma$ in $1.2''$-diameter apertures:
see Bouwens et al.\ 2015) and allow us to exclude any potential lower
redshift contaminants from our samples.

Importantly, our ground-based observations also include
moderately-deep ($\sim$26 mag at $5\sigma$ [1.2$''$-diameter
  apertures]) Y-band observations which we use to constrain the nature
of our selected $z>7$ candidates (where HST observations are
unavailable). These observations are available over the
CANDELS-UDS+COSMOS fields through HAWK-I and VISTA as part of the HUGS
(\citealt{fontana14}) and UltraVISTA (\citealt{mccracken12}) programs,
respectively.  A more detailed description of the
observations we utilize in constructing our source catalogs, as well
as our procedure for constructing these catalogs is provided in
\citet{bouwens15} (see Table~1, Figure~2, and \S3 from
\citealt{bouwens15}).

HST photometry was performed running the Source Extractor software
\citep{bertin96} in dual-image mode, taking the detection images to be
the square root of the $\chi^{2}$ image \citep{szalay99} and
PSF-matching the observations to the H$_{160}$-band PSF. The colors
and total magnitudes were measured with Kron-like (1980) apertures and
Kron factors of 1.6 and 2.5 respectively.

Photometry on sources in the ground-based data is performed after the
contamination from foreground sources is removed, using an automated
cleaning procedure (\citealt{labbe10a}; \citealt{labbe10b}).  The
positions and two-dimensional spatial profiles of the foreground
sources are assumed to match that seen in the high-spatial resolution
HST images, after PSF-matching to the ground-based observations.  The
total flux in each source is then varied to obtain a good match to the
light in the ground-based images.  Light from the foreground sources
is subsequently subtracted from the images, before doing photometry on
the sources of interest.  Flux measurements for individual sources are
then performed in 1.2$''$-diameter circular apertures due to the
objects being inherently unresolved in the ground-based observations.
These flux measurements are then corrected to total, based on the
model flux profiles computed for individual sources based on the
observed PSFs.  The procedure we employ here to derive fluxes is very
similar to that employed in Skelton et al.\ (2014: see also Galametz
et al.\ 2013 and Guo et al.\ 2013 who have also adopted a similar
procedure for their ground-based photometry).

\begin{figure}
\epsscale{1.2}
\plotone{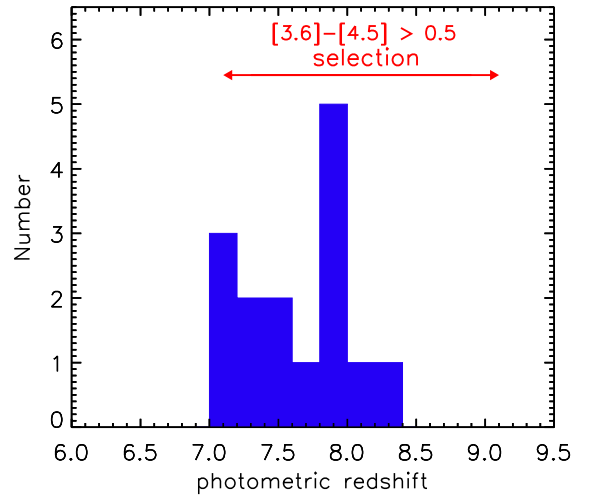}
\caption{The photometric redshift distribution of the $z=7$-9
  $[3.6]-[4.5]>0.5$ IRAC-selected control sample (including 15
  sources) we used to validate our HST+IRAC selection technique (\S3.2).
  The control sample of IRAC-red optical dropouts was identified
  exclusively from those fields with deep $Y$-band data (i.e., CANDELS
  GOODS-North, GOODS-South, UDS, COSMOS, ERS) and consider sources
  that are fainter than we focus on here for our primary selection
  (Table~\ref{tab:table_details}).  Redshift estimates were made,
  based on their observed HST + ground-based photometry.  No
  consideration of the Spitzer/IRAC fluxes is made in deriving the
  photometric redshift presented here (to ensure that the two redshift
  measures are entirely independent).  These results strongly suggest
  that one can use the Spitzer/IRAC [3.6]$-$[4.5] color to reliably
  distinguish $z>7$ galaxies from $z<7$ galaxies (especially in the
  present case where one makes exclusive use of those sources with
  relatively extreme $[3.6]-[4.5]$
  colors).\label{fig:validation_of_method}}
\end{figure}

\subsection{Spitzer/IRAC Data Set and Photometry}

The detailed information we have on $z\sim6$-9 galaxy candidates over
the CANDELS fields from HST is nicely complemented in the mid-IR by
the Spitzer Extended Deep Survey (SEDS, PI: Fazio) program
\citep{ashby13}, which ranges in depth from 12 hours to $>$100 hours
per pointing, though 12 hours is the typical exposure time.  The SEDS
program provides us with flux information at $3.6\mu$m and $4.5\mu$m,
which can be useful for probing $z\sim6$-9 galaxies in the rest-frame
optical, quantifying the flux in various nebular emission lines, and
estimating the redshift.

Over the GOODS-North and GOODS-South fields, we make use of
Spitzer/IRAC reductions, which include essentially all the
Spitzer/IRAC observations obtained to the present (Labb{\'e} et
al.\ 2015: but see also Ashby et al.\ 2015), with 50-200 hours of
observations per pixel in both bands (and typically $\sim$100 hours).

Our procedure for performing photometry on the IRAC data is
essentially identical to that used on the ground-based observations,
except that we utilize $2''$-diameter circular apertures for measuring
fluxes.  These fluxes are then corrected to total based on the model
profile of the individual sources + the PSF.  Depending on the size of
the source, these corrections range from $\sim$2.2$\times$ to
2.4$\times$.

The median $5\sigma$ depths of these Spitzer/IRAC observations for a
$\sim$26-mag source is 25.5 mag in the $3.6\mu$m band and 25.3 mag in
the $4.5\mu$m band.

\section{Sample selection}

\subsection{[3.6]-[4.5] IRAC color vs. redshift and HST detections}
\label{subsec:method}

Many recent studies (e.g., \citealt{schaerer09}; \citealt{shim11};
\citealt{smit14a}; \citealt{labbe13}; \citealt{stark13};
\citealt{debarros14}) have presented convincing evidence to support
the presence of strong nebular line contamination in photometric
filters, particularly for the Spitzer/IRAC [3.6] and [4.5] bands.  The
observed [3.6]$-$[4.5] IRAC color of galaxy candidates appear to be
strongly impacted by the presence of these lines at different
redshifts, in particular those of H$\alpha$ and [OIII].
Figure~\ref{fig:color} provides an illustration of the expected
dependence of the Spitzer/IRAC $[3.6]-[4.5]$ color, as a function of
redshift, assuming an [OIII]+H$\beta$ EW (rest-frame) of $\sim$2250\AA, which is at the high
end of what has been estimated for galaxies at $z\sim7$
(\citealt{labbe13}; \citealt{smit14a}; \citealt{smit14b}).

The significant change in the $[3.6]-[4.5]$ color of galaxies from
$z\sim6$-7 to $z\geq7$ suggests this might be a promising way of
segregating sources by redshift and in particular to identify galaxies
at $z\geq7$.  Such information would be especially useful for search
fields like CANDELS EGS, which lack deep observations in $Y$-band at
$\sim$1.1$\mu$m to estimate the redshifts directly from the position
of the Lyman break.  Smit et al. (2015) have shown that selecting
sources with blue $[3.6]-[4.5]$ colors can effectively single-out
sources at $z\sim6.6$-6.9 over all CANDELS fields, even in the absence
of $Y$-band coverage.

Here we attempt to exploit this strong dependence of the $[3.6]-[4.5]$
color on redshift to identify some of the brightest $z\geq7$ galaxies
over the CANDELS fields. In performing this selection, we start with
the source catalogs derived by Bouwens et al.\ (2015) and Skelton et
al.\ (2014) over a $\sim$900 arcmin$^2$ region from the five CANDELS
fields.  In general, we will rely on the source catalogs from Bouwens
et al.\ (2015) where they exist (covering a 750 arcmin$^2$ area or
$\sim$83\% of CANDELS).\footnote{Bouwens et al.\ (2015) only
  considered those regions in CANDELS where deep optical and near-IR
  observations are available from the CANDELS observations.}
Otherwise, we will rely on the Skelton et al.\ (2014) catalogs and
photometry.

We then apply color criterion to identify a base sample of Lyman-break
galaxies at $z\sim6.3$-9.0.  In particular, over the CANDELS-UDS,
COSMOS, and EGS fields, we use a

\begin{equation}\label{eq:1}
\begin{split}
 & (I_{814}-J_{125} >2.2)\wedge (J_{125}-H_{160}<0.5) \wedge \\
 & (I_{814}-J_{125} > 2(J_{125}-H_{160})+2.2)
\end{split}
\end{equation}

\noindent criterion.  Over the CANDELS-GN and GS fields, we require
that sources satisfy one of the two color criteria defined by
Eq. \ref{eq:2} or Eq. \ref{eq:3}:

\begin{equation}\label{eq:2}
\begin{split}
 & (z_{850}-Y_{105}>0.7)\wedge(J_{125}-H_{160}<0.45)\wedge \\
 & (z_{850}-Y_{105}>0.8(J_{125}-H_{160})+0.7)\wedge \\
 & ((I_{814}-J_{125}>1.0)\vee (SN(I_{814})<1.5))
\end{split}
\end{equation}
 
\begin{equation}\label{eq:3}
\begin{split}
 & (Y_{105}-J_{125}>0.45)\wedge (J_{125}-H_{160}<0.5) \\
 & \wedge(Y_{105}-J_{125}> 0.75(J_{125}-H_{160})+0.525)
\end{split}
\end{equation}

These color criteria are essentially identical to those from
  Bouwens et al.\ (2015) but allow for $J_{125}-H_{160}$ colors as red
  as 0.5 mag to match up with the color criteria of Oesch et
  al.\ (2014) and Bouwens et al.\ (2015) in searching for $z>8.5$
  galaxies (i.e., $J_{125}-H_{160}>0.5$).  In so doing, it was our
  goal to maximize the completeness of our selection for bright
  $z=7$-9 galaxies within the CANDELS program.\footnote{We remark that
    any contaminants in a particularly bright selection would be
    generally easy to identify given the depth of the HST, Spitzer,
    and supporting ground-based observations.}

These color criteria are motivated in Figure 3 of Bouwens et al.\ (2015) and
result in a very similar redshift segregation as one achieves using
photometric redshifts.

We require that sources have $[3.6]-[4.5]$ colors redder than 0.5 mag
(see Figure~\ref{fig:color}).  This color criterion was chosen (1) so
as to require slightly redder colors than was the average color
measured by Labbe et al.\ (2013) for their faint $z\sim8$ sample from
the HUDF (i.e., $\sim$0.4) and (2) such that sources would not easily
satisfy the criterion simply due to noise (requiring $>$2$\sigma$
deviations for the typical source).  To be certain that the IRAC
colors we measured are robust, we exclude any sources where the
subtracted flux from neighboring sources exceeds 65\% of the original
flux in a $2''$-diameter aperture (before subtraction).

To ensure our selection is free of $z<7$ galaxies, we required that
sources show no statistically significant flux at optical wavelengths.
Sources that show at least a $1.5\sigma$ detection in terms of the
inverse-variance-weighted mean $V_{606}$ and $I_{814}$ flux with HST
were excluded.  In addition, we also excluded sources detected at
$>2.5\sigma$ in the deep optical imaging observations available over
each field from the ground.  We adopted a slightly less stringent
threshold for detections in the ground-based observations, due to the
impact of neighboring sources on the overall noise properties.

Finally, we consider potential contamination by low-mass stars,
particularly later $T$ and $Y$ dwarfs (T4 and later), where the
$[3.6]-[4.5]$ color can become quite a bit redder than 0.5 mag
(Kirkpatrick et al.\ 2011; Wilkins et al.\ 2014).  To exclude such
sources from our samples, both the spatial information we had on each
source from the SExtractor stellarity parameter and total SED
information were considered.  Sources with measured stellarities
$>$0.9 were identified as probable stars (where 0 and 1 corresponds to
extended and point-like sources, respectively), as were sources with
measured stellarity parameters $>$0.5 if the flux information we had
available for sources was significantly better fit ($\Delta \chi^2 >
2$) with a low-mass stellar model from the SpecX prism library
(Burgasser et al.\ 2004) than with the best-fit galaxy SED model, as
derived by the Easy and Accurate Zphot from Yale (EAZY;
\citealt{brammer08}) software.  Our SED fits with EAZY considered both
the standard SED templates from EAZY and SED templates from the Galaxy
Evolutionary Synthesis Models (GALEV; \citealt{kotulla09}).  Nebular
emission lines, as described by \citet{anders03}, were added to the
GALEV SED template models assuming a 0.2$Z_{\odot}$ metallicity.  

No sources were removed from our selection as probable low-mass stars.
The procedure we use here to exclude low-mass stars from our selection
is identical to that utilized by \citet{bouwens15}.

\begin{deluxetable*}{cccccccc}
\tablewidth{0pt}
\tablecolumns{8}
\tabletypesize{\footnotesize}
\tablecaption{A complete list of the resulting z$\geq$7 sources
  identified after applying our selection criteria.\label{tab:table_details}}
\tablehead{
\colhead{ID} & 
\colhead{R.A.} & 
\colhead{Dec} &
\colhead{$m_{AB}$\tablenotemark{a}} & 
\colhead{[3.6]$-$[4.5]} & 
\colhead{$z_{phot}$\tablenotemark{b}} & 
\colhead{$Y_{105}-J_{125}$\tablenotemark{c}} & \colhead{References\tablenotemark{*}}}
\startdata
COSY-0237620370 & 10:00:23.76 & 02:20:37.00 & 25.06$\pm$0.06 & 1.03$\pm$0.15 & 7.14$\pm^{0.12}_{0.12}$ & $-$0.13$\pm$0.66 & [1],[2],[3]\\
EGS-zs8-1 & 14:20:34.89 & 53:00:15.35 & 25.03$\pm$0.05 & 0.53$\pm$0.09 & 7.92$\pm^{0.36}_{0.36}$ & 1.00$\pm$0.60 & [3], [4]\\
EGS-zs8-2 & 14:20:12.09 & 53:00:26.97 & 25.12$\pm$0.05 & 0.96$\pm$0.17 & 7.61$\pm^{0.26}_{0.25}$ & 0.66$\pm$0.37 & [3] \\
EGSY-2008532660 & 14:20:08.50 & 52:53:26.60 & 25.26$\pm$0.09 & 0.76$\pm$0.14 & 8.57$_{-0.43}^{+0.22}$ &  
\enddata
\tablenotetext{*}{References: [1] Tilvi et al.\ 2013, [2] Bowler et al.\ 2014, [3] Bouwens et al.\ 2015, [4] Oesch et al.\ 2015b}
\tablenotetext{a}{The apparent magnitude of each source in the $H_{160}$ band.}
\tablenotetext{b}{The photometric redshift estimated by EAZY,
  including flux measurements in the $Y$ band.  The uncertainties quoted here correspond to $1\sigma$.}
\tablenotetext{c}{The $Y-J$ color for each source. The COSMOS candidate
  uses ground based data whilst the EGS candidates use Y$_{105}$ and
  J$_{125}$ filters (where available).}
\end{deluxetable*}

\begin{figure*}
\plotone{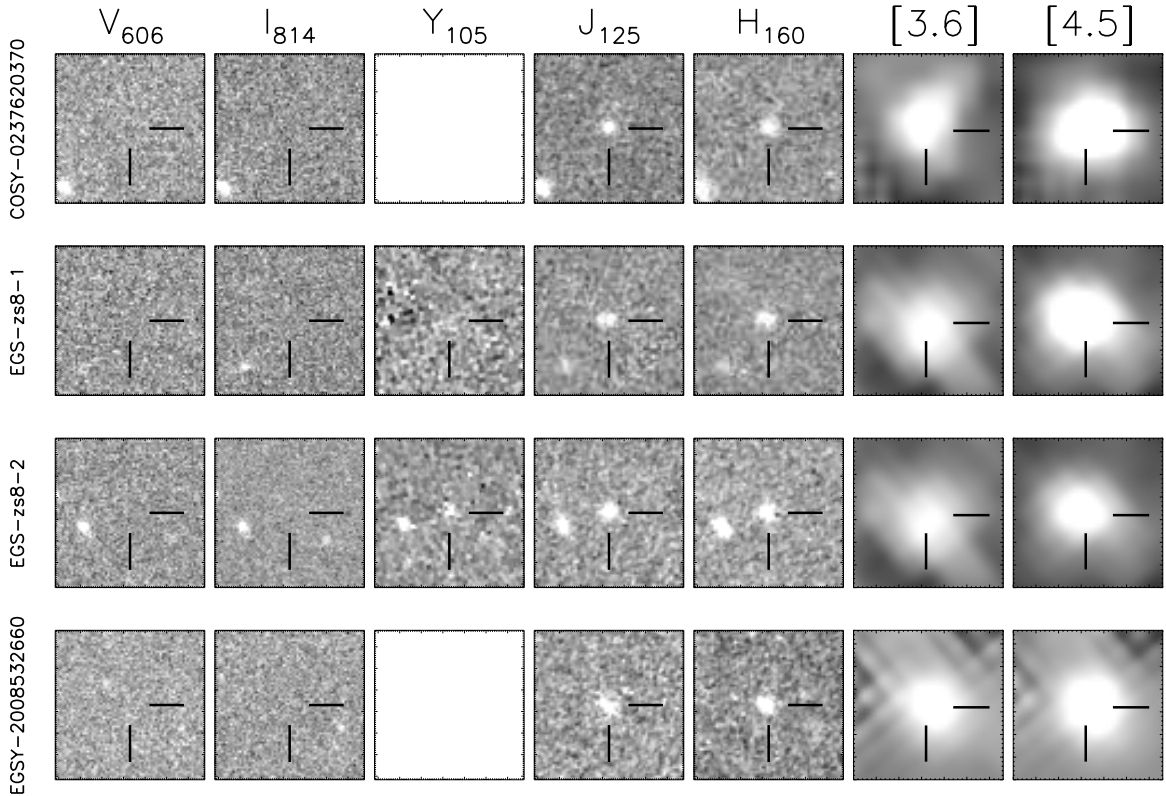}
\caption{HST/ACS $V_{606}I_{814}$, HST/WFC3 $Y_{105}J_{125}H_{160}$,
  and Spitzer/IRAC $3.6\mu$m+$4.5\mu$m postage stamp images
  ($4''\times4''$) of the 3 $z \geq 7$ candidates identified over the
  5 CANDELS fields.  On the Spitzer/IRAC images, flux from
  neighbouring sources has been removed.  $Y$-band observations at
  $1.05\mu$m are also available for COSY-0237620370 from ground-based
  programs (ZFOURGE [Tilvi et al.\ 2013], UltraVISTA [Bowler et
    al.\ 2014]).\label{fig:postage_stamp}}
\end{figure*}

\subsection{Validation of Selection Technique}

Before applying the selection criteria from \S3.1 to the
$\sim$900-arcmin$^2$ CANDELS + ERS search fields, it is useful to
first test these criteria on those data sets which feature deep $z$
and $Y$-band observations.  The availability of observations at these
wavelengths, together with observations at both redder and bluer
wavelengths with HST, allows for very accurate estimates of the
redshifts for individual sources.  There are five data sets that
possess these observations: (1) CANDELS GOODS-S, (2) CANDELS GOODS-N,
(3) ERS, (4) CANDELS UDS, and (5) CANDELS COSMOS field.  The first
three feature these observations with HST and the latter two using
ground-based telescopes.

We apply selection criteria from the previous section to a
$H_{160}$-band limiting magnitude of 26.7 mag for the first three
fields and 26.5 mag for the latter two.  Our decision to use these
depths is partially guided by the sensivity of the Spitzer/IRAC data
over these fields.

Applying the selection criteria from the previous section to the
CANDELS GN+GS and ERS fields ($H_{160,AB}<26.7$), we find 7 sources
that satisfy our selection criteria.  For each of these sources, we
estimate photometric redshifts with EAZY.  In fitting to the observed
photometry, we used the same standard EAZY SED templates as we
described in the previous section.

\begin{figure*}
\epsscale{1.05}
\plotone{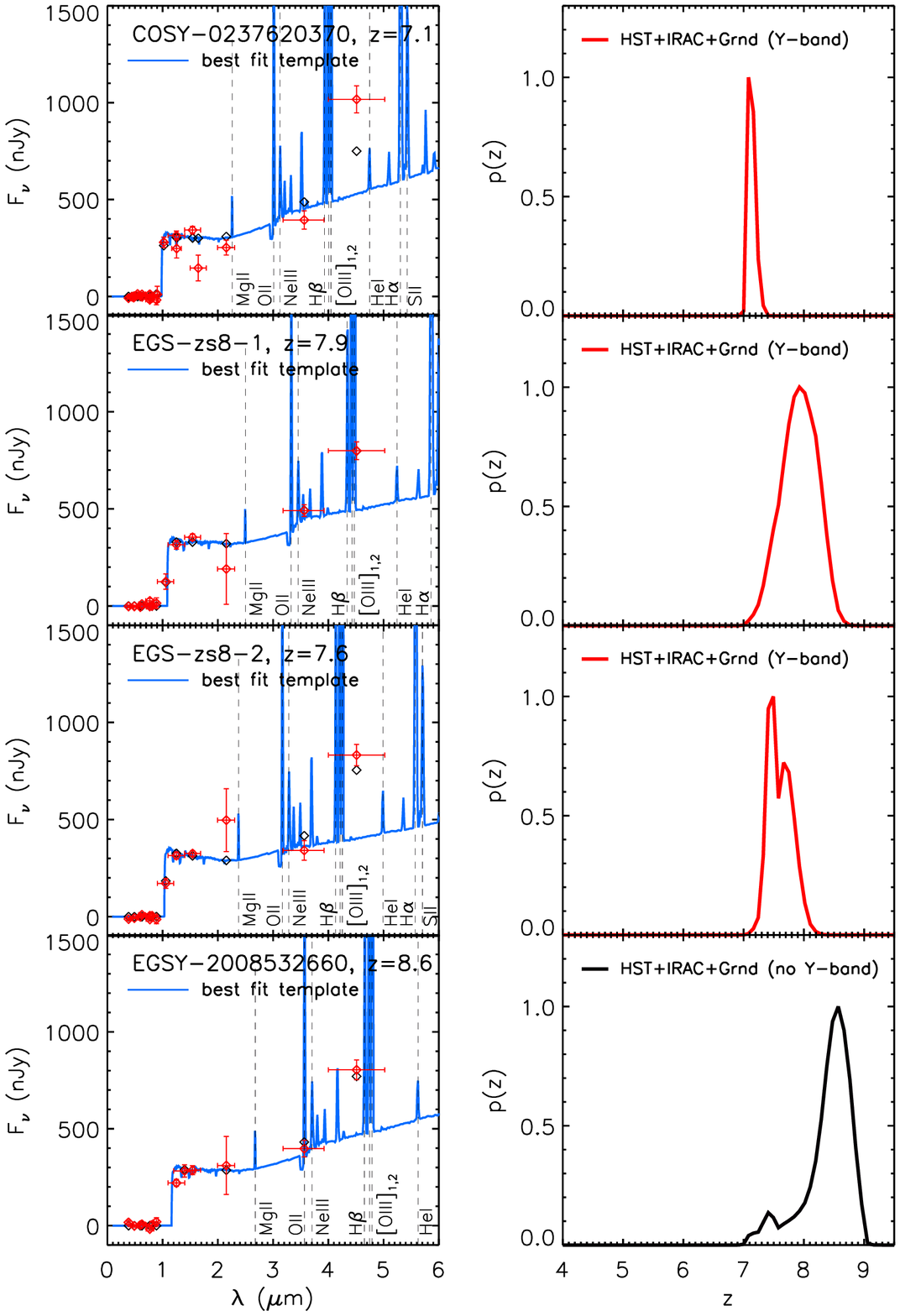}
\caption{\textit{Left}: Best-fit SED models (\textit{blue line}) to
  the observed HST + Spitzer/IRAC + ground-based photometry
  (\textit{red points and error bars}) for the 4 especially bright
  $(H_{160,AB}<25.5$) $z\geq7$ galaxies selected using our IRAC-red
  selection criteria ($[3.6]-[4.5]>0.5$).  Also included on the figure
  is the redshift estimate for the best-fit model SED provided by
  EAZY.  \textit{Right}: Redshift likelihood distributions $P(z)$ for
  the same 4 candidate $z\geq7$ galaxies, as derived by EAZY.  The
  impact of the Spitzer/IRAC photometry on the redshift likelihood
  distributions should be close.\label{fig:sed_pz}}
\end{figure*}

\begin{figure}
\epsscale{0.92}
\plotone{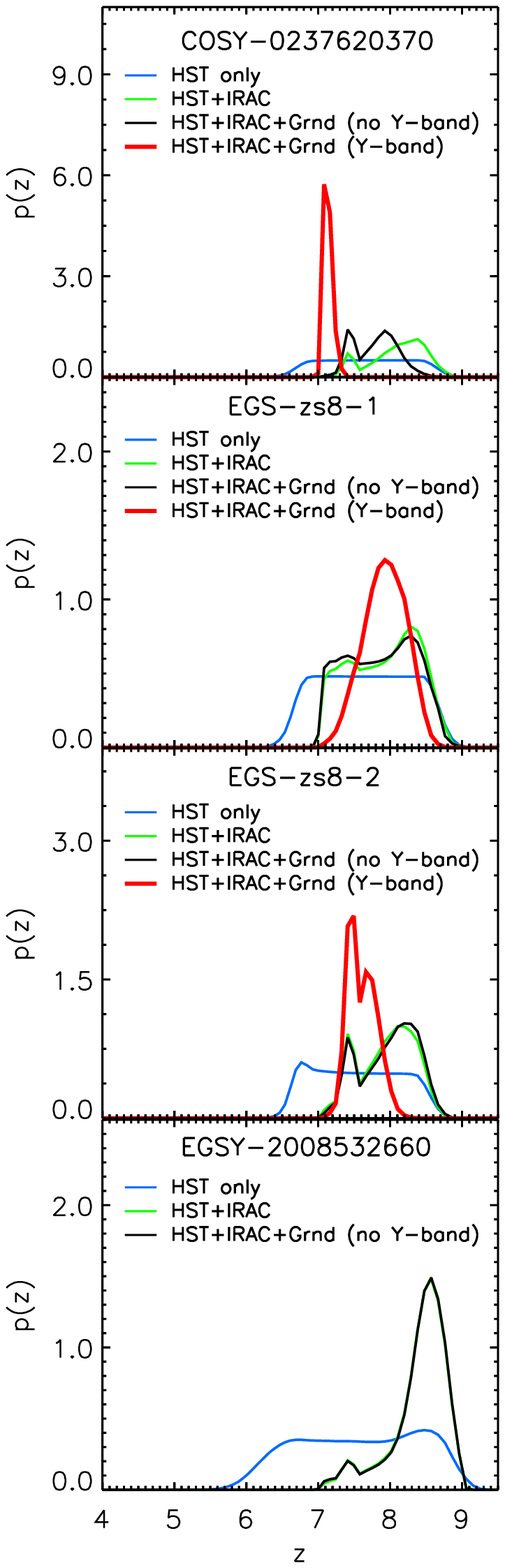}
\caption{Redshift likelihood $P(z)$ constraints for the respective
  galaxy candidates presented in Figure~\ref{fig:sed_pz} shown
  considering the impact of various subsets of the photometry (i.e.,
  HST only, HST+IRAC, HST+IRAC+ground-based observations (no Y-band
  and Y-band included).  It is clear that the addition of both the
  Spitzer/IRAC observations and the $Y$-band observations result in a
  much tighter distribution and allows for a much more accurate
  estimation of the photometric redshift.\label{fig:pz_subset}}
\end{figure}

We also applied the above selection criteria to the CANDELS-UDS and
CANDELS-COSMOS fields, where it is also possible to estimate
photometric redshifts, making use of the available HST observations
and ground-based optical and near-IR $Y$ and $K$ band observations.
Eight sources satisfy these criteria.

All 15 of the sources selected using the criteria from the previous
section are presented in Figure~\ref{fig:validation_of_method} and
fall between $z=7.0$ and $z=8.3$, which is the expected range if a
high-EW [OIII]+H$\beta$ line is responsible for red
$[3.6]-[4.5]$ colors in these galaxies.  This suggests that the
criteria we propose in the previous section can be effective in
identifying a fraction of $z\geq7$ galaxies that are present in fields
with deep HST+Spitzer observations.  The individual coordinates,
colors, and estimated redshifts for individual sources from this
validation sample can be found in Table~\ref{tab:valid_samp} located
in Appendix B.

In recommending the use of the IRAC photometry to subdivide $z\sim6$-9
samples by redshift, we should emphasize that the most robust results
will be obtained making use of only those sources with the smallest
confusion corrections.  While we took care in the selection of both
our primary sample (and the sample we used to validate the technique)
to avoid such souces, such sources were not excluded in making Figure
1 of Smit et al.\ (2015: resulting in a few $z>7$ sources with
anomalously blue Spitzer/IRAC colors).  Despite this issue with Smit
et al.\ (2015) Figure 1, we emphasize that this is nevertheless not a
major concern for sources in their $z=6.6$-6.9 sample.  Only 2 of the
15 sources in the latter sample were subject to a $\sim$3$\times$
correction for flux from neighboring sources and those 2 sources
(GSD-2504846559 and EGS-1350184593) are flagged as less reliable.

\subsection{Search Results for Bright $H_{160,AB}<25.5$ Galaxies}

Here we focus on the identification of only the brightest
$H_{160,AB}<25.5$ $z\geq7$ galaxies using our Spitzer/IRAC color
criteria.  This is to keep the current selection small and to focus on
sources whose surface density was particularly poorly defined from
previous work.  Prior to this work, the only study which identified
such bright $z\sim8$ sources was Bouwens et al.\ (2015).  Focusing on
the brightest sources is also valuable, since it allows us to obtain
very precise constraints on SED shapes and its Spitzer/IRAC colors of
the sources, as well as providing opportunities for follow-up
spectroscopy (see \S4.2).


Applying the selection criteria described in \S\ref{subsec:method} on
the CANDELS-GS, CANDELS-GN, CANDELS-UDS, CANDELS-COSMOS and
CANDELS-EGS fields, we identify a total of 4 especially bright
($H_{160,AB}<25.5$) candidate $z\geq7$ galaxies.  

Our 4 candidate $z\geq$7 galaxies are presented in Table
\ref{tab:table_details} and in Figure \ref{fig:postage_stamp}.  We see
from Figure \ref{fig:postage_stamp} that each candidate is clearly
visible in the HST \textit{H}$_{160}$ and \textit{J}$_{125}$ filters,
as well as the IRAC 3.6 $\mu$m and 4.5 $\mu$m bands.  Of course, no
significant detection is evident in the HST \textit{V}$_{606}$ and
\textit{I}$_{814}$ bands for these sources.  This would suggest that
these sources show a break in their spectrum somewhere between
0.9$\mu$m and $1.2\mu$m and therefore have redshifts between $z\sim6$
and $z\sim8.5$.  [We discuss the impact of information from $Y$-band
  observations available over 3 of the 4 candidates in \S4.1.]

3 of these 4 bright sources are found in the CANDELS-EGS field.
Sources from this field were not included in our earlier attempt to
validate the present selection technique (\S3.2), so only one of these
new sources is in common with the 15 sources just discussed.

To derive constraints on the redshift of each bright source, we again
made use of EAZY.  The photometry provided to EAZY included fluxes
from HST filters, IRAC 3.6 $\mu$m and 4.5 $\mu$m filters and ground
based telescopes.  Using EAZY allows us to generate a best-fit SED of
each galaxy candidate as well as its redshift likelihood distribution
($P(z)$), which we present in Figure \ref{fig:sed_pz}, with the
observed galaxy flux points overplotted.  From the SED plots, we
observe a near-flat rest-frame optical continuum, as well as emission
lines dominating at the location of high flux points, highlighting the
contribution of strong nebular emission lines to the instrument
filters.

One of our $z\geq7$ candidates, i.e., EGS-zs8-2, is sufficiently
compact, as can be seen from Figure~ \ref{fig:postage_stamp}, that we
considered the possibility that it may correspond to a star.  To test
this possibility, we compare its SED to all the stellar SEDs in the
SpecX prism library and find the best-fitting stellar SED.  
The
$\chi^2$ goodness-of-fit for the stellar SED is an order of magnitude
greater than the galaxy SED.  In addition, the SExtractor stellarity
we measure for EGS-zs8-2 in the $J_{125}$ and $H_{160}$ bands is
0.60 and 0.33 (where 0 and 1 correspond to an extended and point
source, respectively), which significantly favors EGS-zs8-2
corresponding an extended source.  Bouwens et al.\ (2015) ran an
extensive number of end-to-end simulations to test the possibility
that point-like sources could scatter to such low measured
stellarities.  Stellarities of $\sim$0.60 are only found for
$H_{160,AB}\sim25$-mag point-like sources in $<$2\% of the simulations
that Bouwens et al.\ (2015) run.  Therefore, both because of the
spatial and spectral information, we can be confident that the
EGS-zs8-2 candidate is a $z\geq7$ galaxy and not a low-mass
star.  However, as we show in \S4.2, perhaps the most convincing piece
of evidence for this source corresponding to a $z>7$ galaxy is our
discovery of a plausible 4.7$\sigma$ Ly$\alpha$ line in the spectrum
of this source at 1.031$\mu$m (Figure~\ref{fig:pascal_lya}).

A second candidate from our selection, COSY-0237620370, is also very
compact and could potentially also correspond to a low-mass star.
However, like EGS-zs8-2, the photometry of the source is better fit
with a galaxy SED than a stellar SED (with a $\chi^2 (star) - \chi^2
(galaxy) = 17.2$) and the source shows evidence for spatial extension,
with a measured stellarity of 0.81 and 0.34 in the $J_{125}$ and
$H_{160}$ bands, respectively.  Stellarities even as high as 0.81 are
only recovered in $\sim$5\% of the end-to-end simulations Bouwens et
al.\ (2015) run at $H_{160,AB}$-band magnitudes of $\sim$25.0.
Earlier, Tilvi et al.\ (2013) obtained the same conclusion regarding
this source based on medium-band observations over this candidate from
the ZFOURGE program, where consistent fluxes are found in the
near-infrared medium bands strongly arguing against this source
corresponding to a low-mass star.  Bowler et al.\ (2014) also conclude
this source is extended and not a low-mass star, based on its spatial
profile (see Figure 6 from Bowler et al.\ 2014) and based on its
observed photometry where $\chi^2 (star) - \chi^2 (galaxy) = 13.0$.

Flux information from HST, Spitzer/IRAC, and ground-based observations
all have value in constraining the redshifts of the candidate $z\geq7$
galaxies we have identified in the present probe.  While the HST flux
information we have available for all three candidates in the
$V_{606}I_{814}J_{125}JH_{140}H_{160}$ bands only allow us to place
them in the redshift interval $z\sim6.5$-9.0 (\textit{blue line} in
Figure~\ref{fig:pz_subset}), we can obtain improved constraints on the
redshifts of the candidates incorporating the flux information from
Spitzer/IRAC and from deep ground-based observations.  Each of these
three candidates appears to have redshifts robustly between $z\sim7.0$
and $z\sim8.6$ (\textit{red and black lines} in
Figure~\ref{fig:pz_subset}).  In addition, as we discuss in \S4.1 and
show in Figure~\ref{fig:pz_subset}, the availability of the
$Y$/$Y_{105}$-band observations allows us to significantly improve our
redshift constraints on all three candidates.

We used the Bouwens et al.\ (2015) catalogs to search 83\% of the
total area of the CANDELS fields and the Skelton et al.\ (2014)
photometric catalogs otherwise (in those regions over the WFC3/IR
CANDELS fields which lack the deep HST/ACS data).  As a check on the
search results we obtained with the Bouwens et al.\ (2015) catalogs,
we applied the same selection criteria to the Skelton et al.\ (2014)
catalogs.  Encouragingly, we identified 75\% of our sample, with only
one candidate missing due to its having a $[3.6]-[4.5]$ color of 0.47
mag.  For all 4 candidates from our primary sample, we find that our
derived [3.6]$-$[4.5] colors are almost identical to those quoted by
Skelton et al.\ (2014), agreeing to $\leq$0.1 mag (and typically
$\Delta$[3.6]$-$[4.5] of 0.05 mag).

We also identified 1 additional bright $(H_{160,AB}<25.5$) $z\geq7$
candidate in the CANDELS-EGS field not identified in our primary
search (see Appendix A).  It seems clear examining its photometry that
this source is extremely likely to be at $z\sim7$-9 (and indeed it
appears in the Bouwens et al.\ 2015 $z\sim8$ sample).  However, since
its measured $[3.6]-[4.5]$ color is 0.22$\pm$0.06 mag in our
photometric catalog (0.3 mag bluer than in the Skelton et al.\ 2014
catalog), we did not include it in our primary sample.  We remark that
photometry for this source was more challenging due to its being
located close to a bright neighbor and its being a two-component
source.


\begin{figure}
\epsscale{1.2}
\plotone{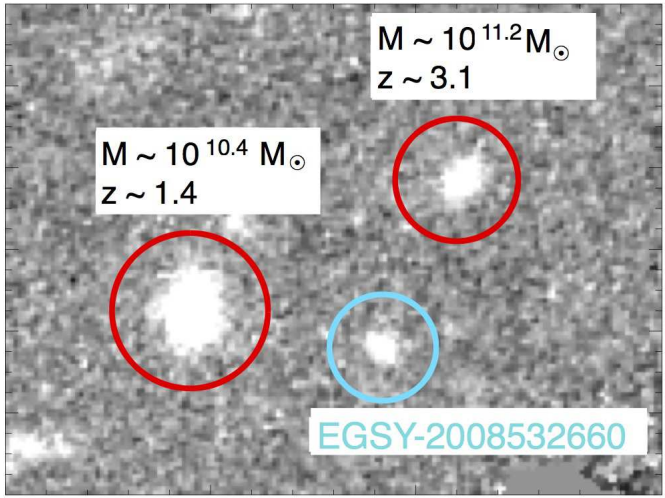}
\caption{Image of the area ($9.6''\times7.2''$) surrounding our new
  $z\sim8.6$ candidate EGSY-2008532660.  EGSY-2008532660 lies very
  close ($<3''$) to two bright, apparently massive foreground galaxies
  (\S3.4).  Based on the position of the foreground sources and their
  inferred masses, we estimate that EGSY-2008532660 is likely
  magnified by a modest factor, i.e.,
  $\sim1.8\times$.\label{fig:lens}}
\end{figure}

\subsection{Possible Evidence for Lensing Amplification of Selected $z>7$ Sources}

For very high redshift sources ($z>>6$), it is expected that the
sources with the brightest apparent magnitudes will benefit from
gravitational lensing (Wyithe et al.\ 2011; Barone-Nugent et
al.\ 2015; Mason et al.\ 2015; Fialkov \& Loeb 2015), and indeed it is
found that a small fraction of the brightest galaxies identified over
the CANDELS program are consistent with being boosted by gravitational
lensing (Barone-Nugent et al.\ 2015).

To investigate whether any of the bright $z\geq7$ galaxies identified
in our search might be gravitationally lensed, we considered all
sources within 5$''$ of our candidates in the Skelton et al.\ (2014)
catalogs and used the estimated redshifts, stellar masses, and sizes
from these catalogs to derive Einstein radii for the foreground
sources assuming a single isothermal sphere model.  We then calculated
the degree to which our bright $z\geq7$ galaxy candidates might be
magnified by the foreground sources.

In only one case was the expected magnification level $>$10\% and this
was for our $z\sim8.6$ candidate EGSY-2008532660.  In this case, we
identified two foreground galaxies which could significantly magnify
this candidate (Figure~\ref{fig:lens}).  The first was a $10^{10.4}$
$M_{\odot}$ mass, $z\sim1.4$ galaxy (14:20:08.81, 52:53:27.2) with a
separation of 2.8$''$ from our $z\sim8.6$ candidate.  The second was a
$10^{11.2}$ $M_{\odot}$ mass, $z\sim3.1$ galaxy (14:20:08.37,
52:53:29.1) with a separation of 2.7$''$ from our $z\sim8.6$
candidate.  Using the measured size of the two sources, we derive
$\sigma\sim170$ km/s and $\sigma\sim370$ km/s for the velocity
dispersion.  We checked and these velocity dispersions are fairly
similar to what fitting formula in Mason et al.\ (2015) yield (i.e.,
using the relation in their Table 1 and applying $H_{160}$-band or
IRAC 3.6$\mu$m apparent magnitudes depending on whether we are
considering the $z\sim1.4$ or $z\sim3.1$ source).

Based on the observed separation of this source from our $z\sim8.6$
galaxy, we estimate a lensing magnification of 20\% and a factor of
1.8 from the former and latter foreground sources.  In computing these
magnification factors, we assume that the mass profile of galaxies is
an isothermal sphere and taking the magnification factor to be
$1/(1-\theta_E/\theta)$ where $\theta$ is separation from the
neighboring sources and $\theta_E$ is the Einstein radius.  Looking at
the morphology of EGSY-2008532660, we see no clear evidence to suggest
that the galaxy is highly magnified and there is no obvious
counterimage.  However, we clearly cannot rule out smaller lensing
amplification factors, particularly if the intrinsic size of the
source is small.  As the inferred stellar or halo masses for the
neighboring galaxies is not precisely known, this translates into a
modest uncertainty into the actual luminosity of this source (as much
as 0.3 dex).  Given this fact, we consider it safest for us to exclude
it from analyses of the $UV$ LF.

\section{Validation of Our $z\sim8$ Selection}

Here we attempt to determine the nature of the $z>7$ candidates we
selected using the HST+Spitzer/IRAC+ground-based observations using
some $Y$-band observations that became available over a few of our
candidates and using the results of some follow-up spectroscopy that
we performed (first reported in Oesch et al.\ 2015b).

\subsection{$Y$-band Photometric Observations}

Deep observations at 1.05$\mu$m are particularly useful in
ascertaining the nature of these candidates and also their redshift,
due to the $Y$-band photometry providing constraints on the position
of the Lyman break as it redshifts from $1.2\mu$m to 0.9$\mu$m.

\begin{figure}
\epsscale{1.2}
\plotone{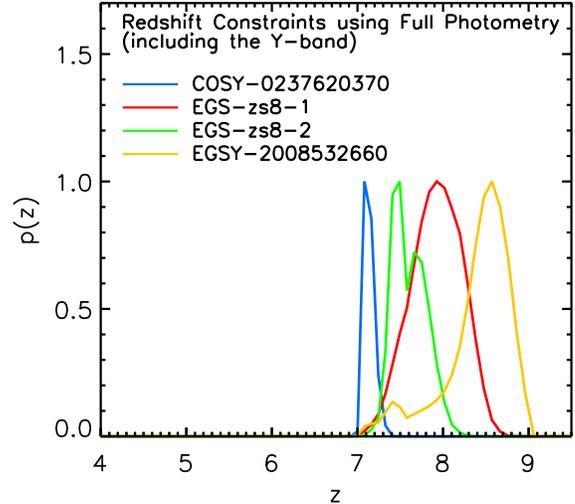}
\caption{Redshift likelihood distributions $P(z)$ for the 4
  $H_{160,AB}<25.5$ sources in our $z\gtrsim 6$, IRAC ultra-red
  selection.  These likelihood distributions include flux constraints
  from the ground-based and HST $Y$-band observations.  Our three
  highest redshift sources have $z_{phot}=7.6\pm0.3$,
  $z_{phot}=7.9\pm0.4$, and
  $z_{phot}=8.6_{-0.4}^{+0.2}$.\label{fig:pz_yband}}
\end{figure}

Deep observations at $1.05\mu$m are available for 3 of the 4 $z\geq7$
candidates that we selected as part of our $H_{160,AB}<25.5$ sample.
$Y$-band observations of the COSY-0237620370 candidate are available
from the 3-year UltraVISTA observations (McCracken et al.\ 2012),
while HST $Y_{105}$-band observations are available over 2 other
candidates in our selection as a result of some recent observations
from the z9-CANDELS follow-up program (Bouwens 2014; Bouwens et
al.\ 2016).\footnote{The purpose of the z9-CANDELS program was to
  determine the nature of high-probability but uncertain candidate
  $z\sim9$-10 galaxies over the CANDELS-UDS, COSMOS, and EGS fields.
  In some cases, bright candidate $z\sim8$ galaxies were located
  nearby bright $z\sim9$-10 candidates and could be readily observed
  in the same pointings.}

\begin{figure*}
\epsscale{0.78}
\plotone{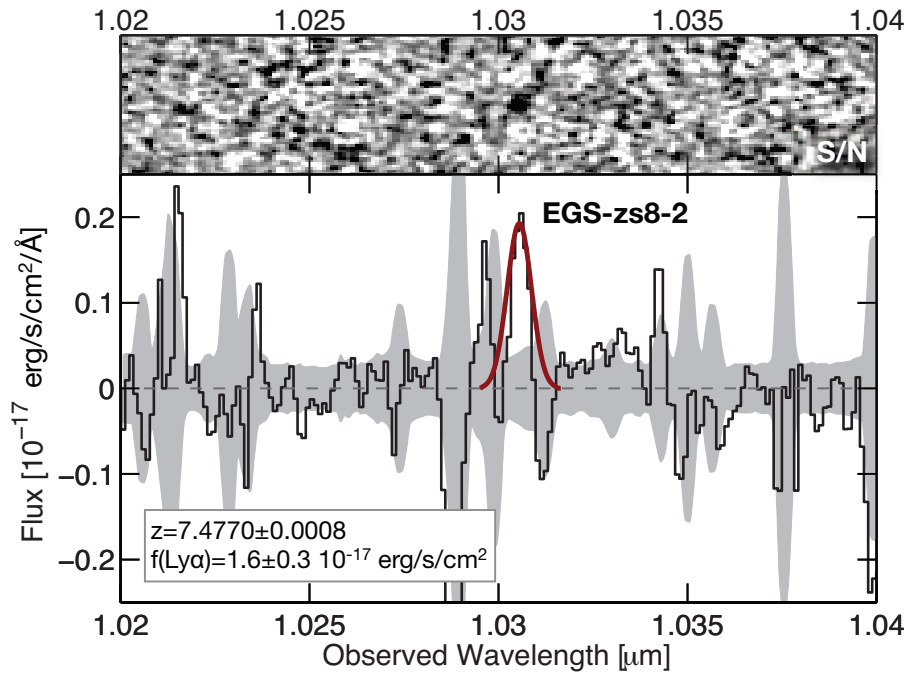}
\caption{Keck/MOSFIRE spectra of EGS-zs8-2.  The 2D spectrum after a
  2x2 binning is presented in the upper panel, while the extracted 1D
  spectrum is shown in the lower panel. The gray shaded area
  represents the 1$\sigma$ flux uncertainty, the red line shows the
  best-fit Gaussian.  A candidate Ly$\alpha$ line (detected at
  4.7$\sigma$ significance) is apparent at 1.031$\mu$m between two sky
  lines.  Using a simple gaussian to model the shape and position of
  this line suggests a redshift of $z=7.4770\pm0.0008$ for this source
  (see also D. Stark et al.\ 2016, in prep).  The other $z>7$
  candidate here targeted with spectroscopy also shows a prominent
  Ly$\alpha$ line, with a measured redshift of $7.7302\pm0.0006$
  (Oesch et al.\ 2015b).\label{fig:pascal_lya}}
\end{figure*}

We make these estimates in an identical way to what we did previously.
Our redshift constraints, including the $Y$-band, are presented in
Table \ref{tab:table_details}.  Furthermore, we present the HST
\textit{Y}$_{105}$ filter images in Figure \ref{fig:postage_stamp},
where we observe a clear detection in the \textit{Y}$_{105}$ filters
for EGS-zs8-1 and EGS-zs8-2 and no detection in the \textit{V}$_{606}$
or \textit{I}$_{814}$ filters, indicating a \textit{z}$\sim$7 Lyman
dropout. For EGS-zs8-1, however, we observe little to no detection in
the \textit{Y}$_{105}$ filter but a clear detection in the
\textit{J}$_{125}$ filter which indicates this galaxy is observed at
\textit{z}$\sim$8.

Figure \ref{fig:sed_pz} and \ref{fig:pz_yband} presents the redshift
likelihood distributions on our $z\geq7$ candidates, incorporating the
$Y$-band observations from UltraVISTA and HST.  It is evident from
Figure~\ref{fig:pz_subset} that the $Y$-band data greatly improves our
constraints on the redshift of the individual candidates in our
selection.  Together with the results in \S3.2 and
Figure~\ref{fig:validation_of_method}, these results largely validate
our selection technique.

\subsection{Keck/MOSFIRE Spectroscopic Follow-up}

\subsubsection{Observations and Reduction}

In addition to using photometric data in the Y-band to validate our
method, we also tested this method by obtaining deep near-IR
spectroscopy on 2 sources from the current selection. Oesch et
al.\ (2015b) already provided a first description of the observational
set-up we utilized for half of our targets, so we keep the current
discussion short.  A total of 4 hours of good Y-band spectroscopy were
obtained in the CANDELS-EGS field with the Multi-Object Spectrometer
for Infra-Red Exploration (MOSFIRE: McLean et al. 2012) instrument on
the Keck I telescope.  Two masks \citep[see Fig. 2 of ][]{oesch15}
were utilized and our spectra were taken with 180 s exposures at a
spectral resolution of R=3500 and R=2850 (for a 0.7'' and 0.9'' slit
respectively) over 3 nights (April 18, April 23, April 25, 2014 -
although due to poor weather conditions, April 18 was effectively
lost), with the aim of searching for Ly$\alpha$ emission in EGS-zs8-1
and EGS-zs8-2.  Each mask contains a slitlet placed on a star, which
we use for monitoring the sky transparency and observing conditions of
each exposure.  These observations were reduced using a modified
version of the DRP MOSFIRE reduction code pipeline (for details see
Oesch et al.\ 2015b).  The spectra complement the photometric data sets
for these two galaxies and allow us to confirm their redshifts.

\subsubsection{Ly$\alpha$ Emission Lines}

The observations carried out with Keck/MOSFIRE revealed candidate
Ly$\alpha$ emission lines in the spectra of both EGS-zs8-1 and
EGS-zs8-2. The detection of a Ly$\alpha$ line for EGS-zs8-1 appears to
be robust (a 6.1$\sigma$ detection with a line flux of
$f_{Ly\alpha}=1.7\pm0.3\times 10^{-17}$ erg s$^{-1}$) and places that
source at $z_{Ly\alpha}$=7.7302$\pm$0.0006, as first reported by Oesch
et al.\ (2015b).\footnote{The flux uncertainties that we derive for
  this candidate and EGS-zs8-2 is almost an order-of-magnitude larger
  than found in observations of similar $z>7$ (e.g., Finkelstein et
  al.\ 2013).  This is in part due to the significantly poorer seeing
  conditions we were subject for the observations (1.00$''$ FWHM
  instead 0.65$''$ for Finkelstein et al.\ 2013).  Another potentially
  significant contributing factor is our relatively conservative
  account of the uncertainties in the line flux measurements,
  including uncertainties that arise from the sky subtraction.  The
  uncertainties we derive are consistent with typical values reported
  by the MOSDEF program (Kriek et al.\ 2015).}

The 1D and 2D spectra for our other targeted $z>7$ candidate EGS-zs8-2
is presented in Figure \ref{fig:pascal_lya} (see also Figure 2 in
Oesch et al.\ 2015b for spectra of the confirmed $z=7.7302\pm0.0006$
candidate).  Using a simple gaussian to determine the central
wavelength of the observed line at 1.031$\mu$m (and ignoring asymmetry
and other effects due to skylines surrounding this candidate
Ly$\alpha$ line), we determine the spectroscopic redshift for the
source to be $z_{Ly\alpha}=7.4770\pm0.0008$, with a detection
significance of 4.7$\sigma$ for the line and a line flux of
$f_{Ly\alpha}=1.6\pm0.3\times 10^{-17}$ erg s$^{-1}$ cm$^{-2}$.  While
this line is only detected at $4.7\sigma$ significance, its reality
appears to be supported by subsequent near-infrared spectroscopy
obtained on this source from independent observing efforts (D. Stark
et al.\ 2016, in prep).

In addition to the Ly$\alpha$-emission lines reported by Oesch et
al.\ (2015b) and this work, Zitrin et al.\ (2015) report the detection
of a 7.5$\sigma$ Ly$\alpha$ line for our EGSY-2008532660 candidate in
new Keck/MOSFIRE observations (June 10-11, 2015). This redshift
measurement sets a new high-redshift distance record for galaxies with
spectroscopic confirmation.  Our photometric selection therefore
contains 3 of the 4 most distant, spectroscopically-confirmed galaxies
to date.

The $z_{Ly\alpha}=7.730$, $z_{Ly\alpha}=7.477$ and
$z_{Ly\alpha}=8.683$ redshifts for EGS-zs8-1, EGS-zs8-2 and
EGSY-2008532660, respectively, are in excellent agreement with the
photometric redshifts derived for these galaxies using HST+IRAC+Ground
based observations and our color criteria. The absolute magnitude and
redshifts of EGS-zs8-1, EGS-zs8-2, and EGSY-2008532660 are presented
in the top panel of Figure~\ref{fig:pascal} in relation to other
$z>6.5$ galaxies with clear redshift determinations from Ly$\alpha$.

The current specroscopy provides us with considerable reassurance that
our proposed color technique is an effective method to search for
bright, $z\geq7$ galaxies.\footnote{Interestingly enough, D. Stark et
  al.\ (2016, in prep) have also spectroscopically confirmed that the
  fourth source (COSY-0237620370) from our sample lies at z=7.15.  As
  such, Ly$\alpha$ emission has been found in all 4 galaxies that make
  up our selection.  Our entire sample has therefore been
  spectroscopically confirmed to lie in the redshift range
  $z=7.1$-9.1, with the spectroscopic redshifts being in excellent
  agreement with our derived photometric redshifts.}

\section{Comparison with Previous Work}

\begin{figure}
\epsscale{1.2} \plotone{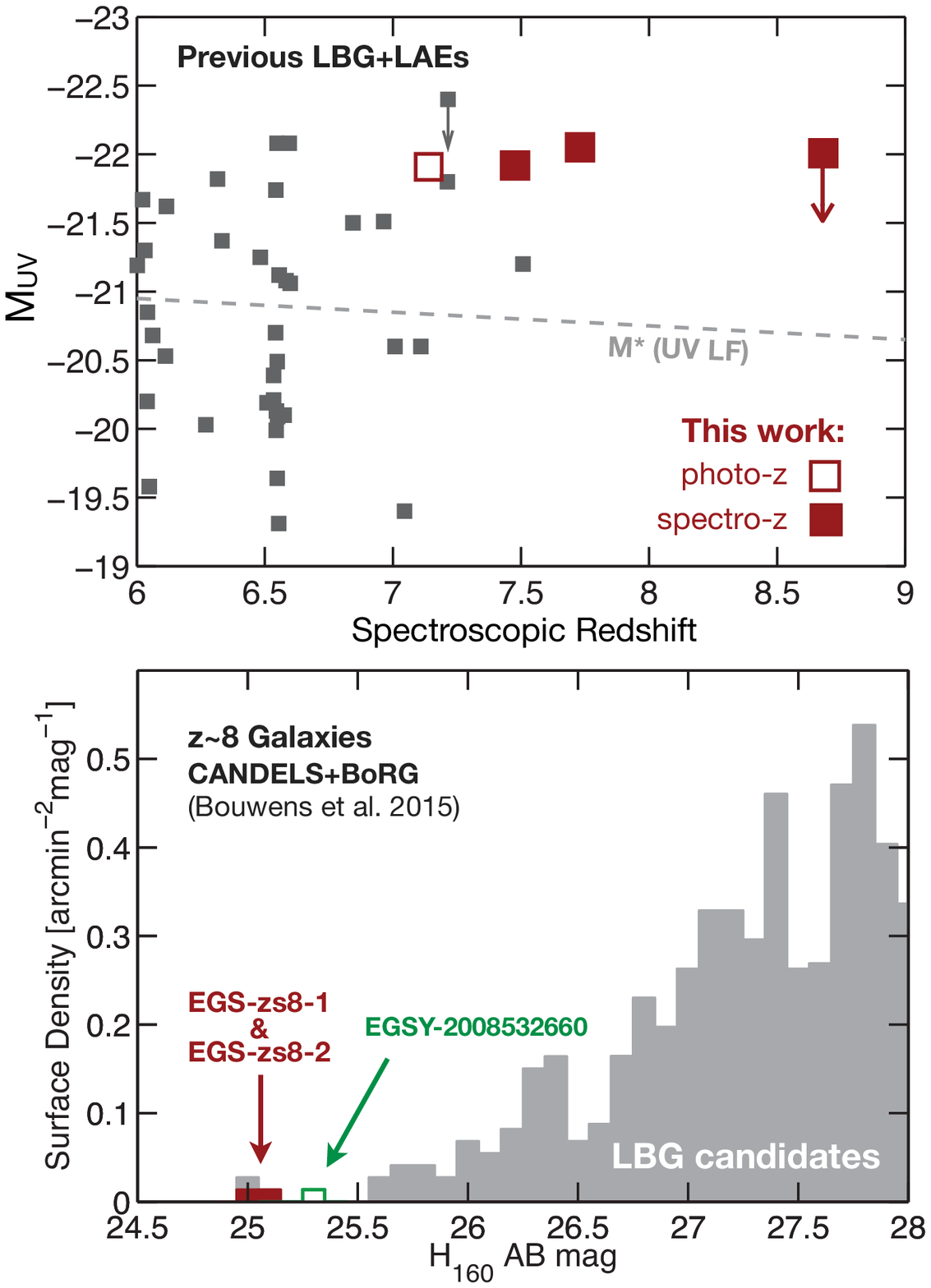}
\caption{(\textit{upper panel}) Absolute magnitudes vs. redshift in
  the rest-frame $UV$ for sources in our current photometric sample
  (solid red squares for the two sources from our sample with redshift
  measurements from spectroscopy and open red squares where the
  redshift estimates derived from the photometry).  For context, the
  absolute magnitudes and redshift measurements for other $z>6$
  galaxies in the literature with spectroscopic redshift measurements
  from Ly$\alpha$ are shown (black squares: compiled from Jiang et
  al. 2013, Finkelstein et al. 2013, Shibuya et al. 2012, Ono et
  al. 2012, and Vanzella et al. 2011).  The gray-dashed line shows the
  evolution of the characteristic magnitude $M^{*}_{UV}$ of the UV LF
  (Bouwens et al. 2015).  Two sources from our sample, with redshift
  measurements from spectroscopy (the 4.7$\sigma$ one requires further
  confirmation) are the brightest $z\gtrsim7.5$ galaxies discovered at
  such high redshifts and similarly for our bright photometric
  $z\sim8.6$ candidate (the downward arrow indicates the likely
  lensing magnification for this candidate: \S3.4).  (\textit{lower
    panel}) Surface density of the full sample of $z\sim$8 galaxies in
  the combined CANDELS and BoRG/HIPPIES fields (Bouwens et al. 2015,
  gray histogram).  The shaded red squares indicate the position of
  our EGS-zs8-1 and EGS-zs8-2 in the Bouwens et al.\ (2015) $z\sim8$
  selections, while the open green square indicates the position of
  the $z\sim8.6$ candidate EGSY-2008532660 (not identified as part of
  the Bouwens et al.\ 2015 $z\sim8$ selection).  Three of the sources
  from our selection represent the brightest-known galaxies at
  $z\gtrsim7.5$ (although one appears to be magnified from
  gravitational lensing).  Interestingly enough, all three of the
  brightest, highest-redshift $z\sim$8 candidates we have identified
  here (and four if one includes the source from Appendix A which is
  also in the Bouwens et al.\ 2015 $z\sim8$ catalog) are located in
  only one of the CANDELS fields (CANDELS EGS), providing an example
  of how dramatic field-to-field variance can be for bright galaxies
  (see also Bouwens et al.\ 2015 and Bowler et
  al.\ 2015).\label{fig:pascal}}
\end{figure}

Three of our four candidates were already identified as part of
previous work.  Tilvi et al.\ (2013) identified COSY-0237620370 as a
$z\sim7$ galaxy by applying Lyman-break-like criteria to the deep
medium-band ZFOURGE data and estimate a redshift of
7.16$_{-0.19}^{+0.35}$.  This source was also identified by Bowler et
al.\ (2014) as a $z\sim7$ galaxy (211127 in the Bowler et al.\ 2014
catalog) using the deep near-IR observations from the Ultra-VISTA
program and derived a photometric redshift of 7.03$_{-0.11}^{+0.12}$
for the source (or 7.20 if the source exhibits prominent Ly$\alpha$
emission), similar to what we find here.  Tilvi et al.\ (2013) derive
a $[3.6]-[4.5]$ color of $1.96\pm0.54$ mag, while Bowler et
al.\ (2014) find 0.7$\pm$0.3 mag, both of which are broadly consistent
with what we find here.

Bouwens et al.\ (2015) identified 3 of the 4 sources as part of their
search for $z\sim7$-8 galaxies over the five CANDELS fields and
segregated the sources into different redshift bins using the
photometric redshift estimates.  The full HST + Subaru Suprime-Cam
$BgVriz$ + CFHT Megacam $ugriyz$ + UltraVISTA $YJHK_s$ photometry was
used to estimate these redshifts for the candidate in the COSMOS
field.  Meanwhile, the HST + CFHT Megacam $ugriyz$ + WIRCam $K_s$ +
Spitzer/IRAC $3.6\mu$m$4.5\mu$m photometry was used in the case of the
two EGS candidates.

Bouwens et al.\ (2015) derived a photometric redshift of $z=7.00$ for
COSY-0237620370 over the CANDELS-COSMOS field and derived photometric
redshifts of 8.1 for the two sources over the CANDELS-EGS field
(EGS-zs8-1 and EGS-zs8-2, respectively), so the latter two
candidates were placed in the $z\sim8$ sample of Bouwens et
al.\ (2015).

There was, however, some uncertainty as to both the robustness and
also the precise redshifts of the CANDELS-EGS candidates from Bouwens
et al.\ (2015).  Prior to the present study, the use of the
$[3.6]-[4.5]$ color has never been systematically demonstrated to work
for the identification of galaxies with redshifts of $z>7$ despite
there being $\sim$5 prominent examples of $z\geq7$ galaxies with
particularly red $[3.6]-[4.5]$ colors (Bradley et al.\ 2008; Ono et
al.\ 2012; Finkelstein et al.\ 2013; Tilvi et al.\ 2013; Laporte et
al.\ 2014, 2015).  Moreover, no $Y_{105}$-band observations were
available over either $z\geq7$ candidate from the CANDELS-EGS field in
the Bouwens et al.\ (2015) selection to validate potential $z\geq7$
galaxies (though such observations have fortuitously become available
as a result of observations made from the z9-CANDELS follow-up program
[Bouwens et al.\ 2016]).

The apparent magnitudes of the $z=7.1$-8.5 galaxies identified as part
of the current selection are much brighter than the typical galaxy at
$z\sim8$, as is evident in both the upper and lower panels in
Figure~\ref{fig:pascal}.  In fact, 3 of the sources from our current
IRAC-red $[3.6]-[4.5]>0.5$ selection appear to represent the brightest
$z\gtrsim7.5$ galaxies known in the entire CANDELS program and
constitute 3 of the 4 $z\sim8$ candidates shown in the lower panel of
Figure~\ref{fig:pascal}.  The only other especially bright
$H_{160,AB}\sim25.0$ $z\sim8$ candidate shown in that lower panel is
presented in the appendix (since it satisfies our $[3.6]-[4.5]>0.5$
selection criteria using an independent set of photometry, i.e.,
Skelton et al.\ 2014).

Interestingly enough, all 4 of the brightest candidates shown in the
lower panel of Figure~\ref{fig:pascal} are located in the CANDELS EGS
field, providing a dramatic example of how substantial field-to-field
variations in the surface densities of bright sources might be (though
we note that EGSY-2008523660 is likely gravitationally lensed).  This
seems to be just a chance occurrence, as none of these candidates is
clearly in a similar redshift window.  The probability that the 4
brightest $z\sim8$ sources in the CANDELS program would be found in
the same CANDELS field (even if one is gravitationally lensed) is
$\sim$1\%.\footnote{There is only one source from our combined
  $z\sim8$ selection with Bouwens et al.\ (2015) which would be
  potentially easier to select as a $z\sim8$ galaxy over the CANDELS
  EGS field.  It is presented in Appendix A.  Its redshift is not well
  constrained (lying anywhere between $z\sim 7.1$ and 8.5), but it
  would be marginally easier to find over the CANDELS EGS field since
  the Bouwens et al.\ (2015) $z\sim8$ sample extends down to $z\sim7$
  over that field while the Bouwens et al.\ (2015) $z\sim8$ samples
  over the other fields only extend down to $z\sim7.3$.}

Previously, this point had been strongly made by Bouwens et
al.\ (2015) in discussing the number of bright sources over the
different CANDELS fields (Figure~14, Appendices E and F from Bouwens
et al.\ 2015) and also quite strikingly by Bowler et al.\ (2015) in
comparing the number of bright $z\sim6$ galaxies over the UltraVISTA
and UDS fields.

\section{Implications for the Bright End of the $z\sim8$ LF}

In this section, we will examine the implications of our present
search results for the volume density of luminous galaxies in the
$z\sim7$-9 Universe.  First, we estimate how complete we might expect
our selection to be based on the $[3.6]-[4.5]$ color distribution in
fields where the redshift can be constrained using deep $Y$-band data
(\S6.1).  Second, we make use of our search results and our
completeness estimates to set a constraint on the bright end of the
$z\sim7$-9 LF (\S6.2).

\begin{deluxetable*}{ccccccc}
\tablewidth{0pt}
\tablecolumns{6}
\tabletypesize{\footnotesize}
\tablecaption{Brightest $z\gtrsim 7.5$ Galaxies over the CANDELS Fields and
  $\sim$200 arcmin$^2$ BoRG/HIPPIES area searched by Bouwens et
  al.\ (2015)\label{tab:brightz8}}
\tablehead{
\colhead{ID} & 
\colhead{R.A.} & 
\colhead{Dec} &
\colhead{$m_{AB}$\tablenotemark{a}} & 
\colhead{$z_{phot}$\tablenotemark{b}} & \colhead{$[3.6]-[4.5]$} & \colhead{References\tablenotemark{*}}}
\startdata
EGS-zs8-1 & 14:20:34.89 & 53:00:15.35 & 25.03$\pm$0.05 & 7.9$\pm$0.4 & 0.53$\pm$0.09 & [1], [8]\\
                & & & & ($z_{spec}$=7.7302$\pm$0.0006) \\
EGS-zs8-2 & 14:20:12.09 & 53:00:26.97 & 25.12$\pm$0.05 & $7.6\pm0.3$ & 0.96$\pm$0.17 & [1]\\
                & & & & ($z_{spec}$=7.4770$\pm$0.0008) \\
EGSY-2008532660\tablenotemark{$\dagger$} & 14:20:08.50 & 52:53:26.60 & 25.26$\pm$0.09\tablenotemark{$\ddagger$} & 8.57$_{-0.43}^{+0.22}$\tablenotemark{$\dagger$} & 0.76$\pm$0.14 & \\
                & & & & ($z_{spec}$=8.683$_{-0.004}^{+0.001}$) \\
GNDY-6379018085 & 12:36:37.90 & 62:18:08.50 & 25.44$\pm$0.04 & 7.508\tablenotemark{c} & 0.88$\pm$0.11 & [6]\\
BORGY-9469443552 & 04:39:46.94 & $-$52:43:55.20 & 25.56$\pm$0.20 & 8.29$_{-1.01}^{+0.34}$ & --- & [1],[2],[3], [7]\\
GSDY-2499348180 & 03:32:49.93 & $-$27:48:18.00 & 25.58$\pm$0.05 & 7.84$_{-0.29}^{+0.15}$ & 0.08$\pm$0.09 & [1],[3],[4],[5]\\
BORGY-6504943342 & 14:36:50.49 & 50:43:34.20 & 25.69$\pm$0.08 & 7.49$_{-3.17}^{+0.13}$ & --- & [1]\\
COSY-0235624462 & 10:00:23.56 & 02:24:46.20 & 25.69$\pm$0.07 & 7.84$_{-0.18}^{+0.37}$ & 0.88$\pm$0.61 & [1]\\
BORGY-2463351294 & 22:02:46.33 & 18:51:29.40 & 25.78$\pm$0.15 & 7.93$_{-0.21}^{+0.59}$ & --- & [1]\\
BORGY-2447150300 & 10:32:44.71 & 50:50:30.00 & 25.91$\pm$0.20 & 7.93$_{-0.19}^{+0.48}$ & --- & [1]\\
BORGY-5550543040 & 07:55:55.05 & 30:43:04.00 & 25.98$\pm$0.21 & 7.66$_{-5.63}^{+0.82}$ & --- & [1]\\
Median & & & & & 0.82$_{-0.20}^{+0.07}$ &  \\
\\
\multicolumn{7}{c}{Other Possible Bright $z>7.5$ Galaxies over CANDELS (Appendix A)}\\
EGSY-9597563148 & 14:19:59.75 & 52:56:31.40 & 25.03$\pm$0.10 & 8.19 & 0.22$\pm$0.06 & [1]
\enddata
\tablenotetext{*}{References: [1] Bouwens et al.\ 2015, [2] Bradley et al.\ 2012, [3] McLure et al.\ 2013, [4] Yan et al.\ 2012, [5] Oesch et al.\ 2012, [6] Finkelstein et al.\ 2013, [7] Schmidt et al.\ 2014, [8] Oesch et al.\ 2015b.}
\tablenotetext{a}{The apparent magnitude of each source in the $H_{160}$ band.}
\tablenotetext{b}{Maximum likelihood photometric redshift estimate from EAZY.}
\tablenotetext{c}{Spectroscopic redshift determination (Finkelstein et al.\ 2013).  Finkelstein et al.\ (2013) report a $[3.6]-[4.5]$ color of 0.98$\pm$0.14.}
\tablenotetext{$\dagger$}{Photometric redshift estimate is not based on
  deep $Y$-band imaging observations over candidate of red
  $[3.6]-[4.5]$ color as derived from the Bouwens et al.\ (2015)
  photometry.  Inclusion in this list is based upon a red
  $[3.6]-[4.5]$ color as derived from the Skelton et al.\ (2014)
  photometry.}
\tablenotetext{$\ddagger$}{The flux of this source seems likely to be boosted  by gravitational lensing (\S3.4).}
\end{deluxetable*}

\subsection{[3.6]-[4.5] Color Distribution of $z>7$ Galaxies and the Implications
for the Completeness of our red IRAC Criteria and the [OIII]+H$\beta$ EWs}

In our attempts to identify bright $z>7$ galaxies, we only consider
those sources with red $[3.6]-[4.5]>0.5$ Spitzer/IRAC colors to ensure
that the sources we select are robustly at $z>7$ (see \S3.2).
However, by making this requirement, we potentially exclude those
$z>7$ galaxies which have bluer [3.6]-[4.5] colors, either due to
lower-EW [OIII]+H$\beta$ lines or simply as a result of noise in the
photometry.

To determine how important this effect is, we look at the [3.6]-[4.5]
color distribution of galaxies which we can robustly place at a
redshift $z>7$ (where both lines in [OIII] doublet fall in the [4.5]
band).  The most relevant sources are those bright galaxies we can
place at $z>7$ based on the available HST+ground-based photometry and
which include deep flux measurements at $1\mu$m.  Such measurements
are available for the CANDELS GOODS-S, GOODS-N, UDS, and COSMOS
fields, and a small fraction of the CANDELS EGS field.

For our fiducial results here, we only consider selected sources from
those fields brightward of $H_{160,AB}=26$ and with redshift estimates
greater than $z\gtrsim7.5$.  This is to ensure that we only include
bona-fide $z=7.1$-9.1 galaxies (where the [OIII]+H$\beta$ line in the
4.5$\mu$m band) in our selection.  Photometric redshift errors often
have an approximate size of $\Delta$z$\sim$0.3 to this magnitude
limit, and so to avoid $z<7$ sources scattering into our selection, we
kept our cuts fairly conservative.

The list of such sources at such bright magnitudes is still somewhat
limited at present, with only the bright $\sim$25.6-mag galaxy in the
CANDELS South from Yan et al.\ (2012) and Oesch et al.\ (2012), a
bright $25.7$-mag galaxy in the CANDELS COSMOS field from Bouwens et
al.\ (2015), a bright $\sim$25.5 mag source over the CANDELS
GOODS-North field by Finkelstein et al.\ (2013), two bright sources
over the CANDELS EGS field where $Y_{105}$-band photometry is
available (EGS-zs8-1, EGS-zs8-2), and a third bright source over the
CANDELS EGS field where the $J_{125}-H_{160}$ color allows us to place
it at $z>8$ (EGSY-2008532660).

\begin{figure}
\epsscale{1.2}
\plotone{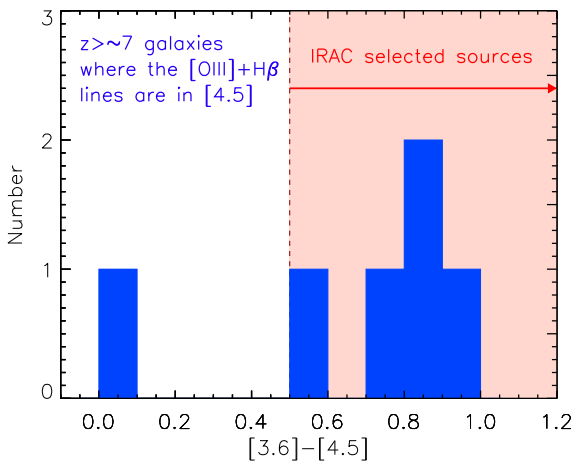}
\caption{Range of Spitzer/IRAC $[3.6]-[4.5]$ colors (blue histogram)
  observed for bright $z>7$ galaxies where we are confident that the
  [OIII]+H$\beta$ emission line falls in the [4.5] band (\S6.1).
  Sources are included in this histogram if they are brighter than
  $H_{160,AB}=26$ and the redshift information available on these
  sources confidently place them at $z\gtrsim7.5$.  This color
  distribution is compared against the IRAC red $[3.6]-[4.5]>0.5$
  selection criteria we use (shaded red region).  The sources
  presented here are the same as those sources in
  Table~\ref{tab:brightz8} with IRAC color measurements.  Five out of
  6 $z\gtrsim7.5$ galaxies show Spitzer/IRAC colors redder than 0.5.
  At face value, this suggests that our proposed $[3.6]-[4.5]>0.5$
  selection would identify 83$_{-16}^{+11}$\% of the bright $z>7$
  population (but we emphasize this percentage is very uncertain due
  to the small numbers involved).  The observed $[3.6]-[4.5]$ color
  distribution also implies a minimum EW for [OIII]+H$\beta$ of
  1300\AA.\label{fig:histo_data}}
\end{figure}

Of these sources, five out of the six sources have [3.6]$-$[4.5]
colors in excess of 0.5, and therefore for simplicity, we will assume
that our IRAC red selection is 83\% complete, but we emphasize that
the completeness correction we derive from this selection is uncertain
and could be much larger (as indeed one would expect if the
[3.6]$-$[4.5] color measurement derived by Labb{\'e} et al.\ 2013,
i.e., $\sim$0.4 mag, for the average stacked $z\sim8$ galaxy are
indicative).

To investigate this possibility, we examined a slightly larger sample
of objects over the four fields where we have photometric redshifts
using $Y$-band imaging.  Considering sources to a $H_{160}$-band
magnitude limit of 26.2 over the CANDELS-UDS and COSMOS fields and
26.7 over the CANDELS GOODS-North and GOODS-South while extending the
photometric redshift selection to $z>7.3$, 6 out of 9 sources satisfy
the [3.6]$-$[4.5]$>$0.5 criterion.  While this suggests the actual
fraction of $z>7$ galaxies with such red IRAC colors may be less than
83\%, this fainter sample is still consistent with our fiducial
percentage.  It also reassuring that our suggested selection criteria
would also apply to GN-108036, the bright $JH_{140}=25.17$ $z=7.213$
galaxy found by Ono et al.\ (2012), given its measured [3.6]$-$[4.5]
color of 0.58$\pm$0.18 mag.

We include a list of those sources and other sources in
Table~\ref{tab:brightz8} from the Bouwens et al.\ (2015) catalog and
also including the bright $z=7.508$ galaxy from Finkelstein et
al.\ (2013).  In Figure~\ref{fig:histo_data}, we present the
[3.6]-[4.5] color distribution for the brightest sources we know to
robustly lie at $z\gtrsim7.5$ based on spectroscopy or from the
available HST+ground-based photometry for those sources that lie in
regions of CANDELS with $Y$-band observations or with
$J_{125}-H_{160}$ colors red enough to confidently place the sources
at $z>8$.

The median [3.6]$-$[4.5] color that we measure is
0.82$_{-0.20}^{+0.08}$ mag.  Such a color implies a minimum EW of
$\sim$1300\AA$\,$ for the [OIII]+H$\beta$ lines, assuming a flat
stellar continuum and no line contribution to the [3.6] band.
However, we emphasize that if there is also a substantial line
contribution to the [3.6] band, e.g., from H$\gamma$, H$\delta$, and
[OII], then the implied EW of the [OIII]+H$\beta$ lines would be much
larger.  For example, adopting the line ratios from the Anders \&
Fritze-v.~Alvensleben (2003), 0.2$\,Z_{\odot}$ model would imply an EW
of $\gtrsim$2100\AA$\,\,$for the [OIII]+H$\beta$ lines.

Some correction is required to the median [3.6]$-$[4.5] color
measurement to account for the fact that the $z>7$ sources from the
CANDELS EGS field were explicitly selected because of their red IRAC
colors.  If we assume that the intrinsic [3.6]$-$[4.5] color for
sources over the CANDELS EGS field is $\sim$0.6 mag (which is the
value we find from the 3 candidates over the other CANDELS fields: see
Table~\ref{tab:brightz8}) and the noise + scatter is $\sim$0.4 (the
value from the other fields), we compute a bias of 0.24 mag from a
simple Monte-Carlo simulation.  Accounting for such biases reduces the
median [3.6]-[4.5] color of the population by 0.24 mag, which implies
a median EW of $\sim$800\AA$\,$and 1500\AA, ignoring and accounting
for a possible nebular contribution to the [3.6] band respectively.

\begin{figure*}
\epsscale{0.78} \plotone{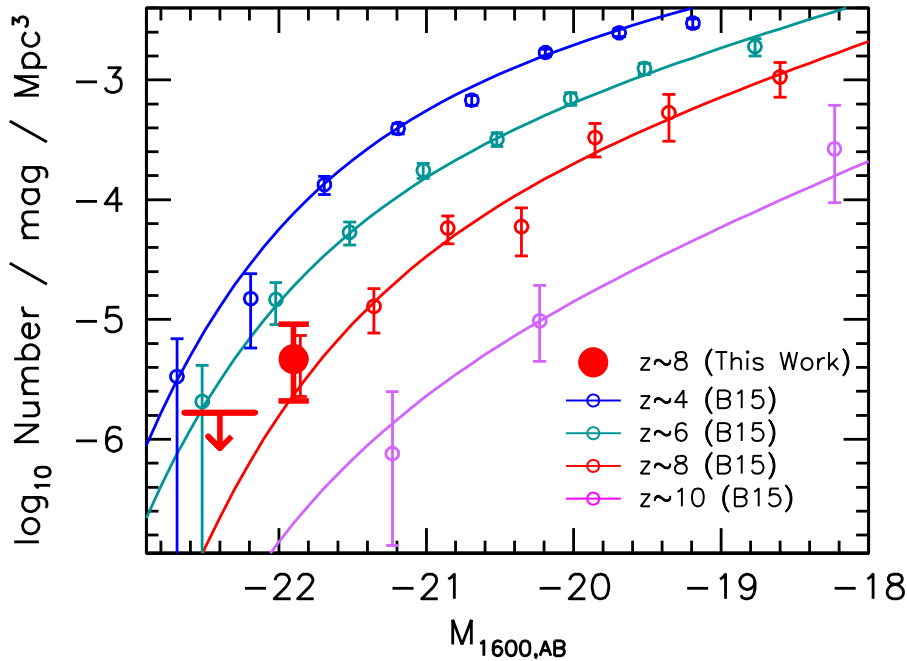}
\caption{Estimated volume density of especially luminous
  $z\sim7.1$-9.1 galaxies using the current IRAC-red search results
  over the entire CANDELS program (\S6.2).  This selection is assumed
  to be 17$_{-11}^{+16}$\% incomplete based on the results from \S6.1
  and excludes the $z\sim8.6$ candidate galaxy EGSY-2008532660 due to
  evidence that this galaxy may be magnified by a foreground source
  (\S3.4).  For context, the $UV$ luminosity function results of
  Bouwens et al. (2015: B15) at $z\sim4$, $z\sim6$, $z\sim8$, and
  $z\sim10$ are also shown with the blue, cyan, red, and magenta open
  circles and solid lines, respectively.  The present constraints on
  the $UV$ LF are consistent with those previously derived by Bouwens
  et al.\ (2015).
\label{fig:lf8}}
\end{figure*}

\subsection{Volume Density of Bright $z\sim8$ Galaxies}

Here we use the search results from the previous section to set a
constraint on the bright end of the $z\sim8$ LF.  We begin this
section by calculating the total selection volume in which we would
expect to find bright $z\geq7$ galaxies with our selection criteria.


We will estimate the selection volumes in a similar way to the
methodology used by Bouwens et al.\ (2015) in deriving the LFs from
the full CANDELS program.  In short, we create mock catalogs over each
search field, with sources distributed over a range in both redshift
$z\sim6$-10 and apparent magnitude ($H_{160,AB}=24$ to 26).  We then
take the two-dimensional $i_{775}$-band images of similar luminosity,
randomly-selected $z\sim4$ galaxies from the HUDF (Bouwens et
al. 2007, 2011, 2015) and create mock images of the sources at
higher-redshift using the two dimensional pixel-by-pixel $z\sim4$
galaxies as a guide (see Bouwens et al. 1998, 2003), adopting random
orientations relative to their orientation in the HUDF and scaling
their physical sizes as $(1+z)^{-1.2}$, which is the approximate
relationship that has been found comparing the mean size of galaxies
at fixed luminosity, as a function of redshift (Oesch et al.\ 2010;
Grazian et al.\ 2012; Ono et al.\ 2013; Holwerda et al.\ 2015;
Kawamata et al.\ 2015; Shibuya et al.\ 2015).  Individual sources were
assigned $UV$ colors based on their $UV$ luminosity, using the
$\beta$-$M_{UV}$ relationship derived by Bouwens et al.\ (2014) and
allowing for an intrinsic scatter $\sigma_{\beta}$ of 0.35 at high
luminosities ($M_{UV,AB}=[-22,-20]$) as found by Bouwens et
al.\ (2009, 2012) and systematically decreasing to 0.15 at lower
luminosities as found by Rogers et al.\ (2014).

In addition to the HST images we created for individual sources, we
also constructed simulated ground-based and Spitzer/IRAC images for
these sources which we added to the real ground-based + Spitzer/IRAC
data.  These simulated images were generated based on the mock
$H_{160}$-band images we constructed for individual sources and
convolving by the $H_{160}$-to-IRAC, $H_{160}$-to-ground kernels that
\textsc{Mophongo} (Labb{\'e} et al.\ 2010) derived from the
observations.  In producing simulated IRAC images for the mock
sources, we assume a rest-frame EW of 300\AA$\,$for H$\alpha$+[NII]
emission and 500\AA$\,$for [OIII]+H$\beta$ emission over the entire
range $z=4$-9, a flat rest-frame optical color, and a
$H_{160}$-optical continuum color of 0.2-0.3 mag, to match the
observational results of Shim et al.\ (2011), Stark et al.\ (2013),
Gonz{\'a}lez et al.\ (2012, 2014), Labb{\'e} et al.\ (2013), Smit et
al.\ (2014, 2015), and Oesch et al.\ (2013).

We took the simulated images we created for individual sources and
added them to the real HST, ground-based, and Spitzer/IRAC
observations.  These simulated images were, in turn, used to construct
catalogs and our selection criteria applied to the derived catalogs in
exactly the same way as we applied these criteria to the real
observations (including excluding sources which violated our confusion
criteria).

Summing the results over all five CANDELS fields, we compute a total
selection volume of 1.6$\times$10$^{6}$ Mpc$^{3}$ per 1-mag interval
for galaxies with $H_{160,AB}$ magnitudes brightward of 25.5.  If we
assume that the present selection of $z\sim8$ galaxies is complete,
this would imply a volume density of $<$1.4$\times$10$^{-6}$
Mpc$^{-3}$ mag$^{-1}$ and $3.8_{-2.1}^{+3.7}$$\times$10$^{-6}$
Mpc$^{-3}$ mag$^{-1}$ for $H_{160,AB}\sim24.5$-25.0 and
$H_{160,AB}\sim25.0$-25.5 galaxies, respectively.  We ignore the
contribution of the $z\sim8.6$ candidate galaxy EGSY-2008532660, given
the evidence that it may be slightly magnified (\S3.4).

However, we cannot assume that the present selection of bright
$z\sim8$ galaxies is complete, since not every $z\sim8$ galaxy
exhibits such a red $[3.6]-[4.5]$ color.  In the previous section, we
found that only 5 out of the 6 bright ($H_{160,AB}<26$), secure
$z\gtrsim7$ sources within CANDELS showed such red galaxies.
Correcting the volume densities given in the previous paragraph to
account for this slight empirically-derived incompleteness
(0.83$_{-0.16}^{+0.11}$), we estimate volume densities of
$<$1.7$\times$10$^{-6}$ Mpc$^{-3}$ mag$^{-1}$ and
$4.7_{-2.7}^{+4.6}$$\times$10$^{-6}$ Mpc$^{-3}$ mag$^{-1}$ for
$H_{160,AB}\sim24.5$-25.0 and $H_{160,AB}\sim25.0$-25.5 galaxies,
respectively.  The uncertainty in the completeness estimate is
included in the error we quote for  the volume density.

It is interesting to compare these constraints on the volume density
of bright $z\sim7.1$-8.5 galaxies with other recent constraints which
are available on the volume density of bright end of the $UV$ LFs at
$z\sim4$, $z\sim6$, $z\sim8$, and $z\sim10$ galaxies from
state-of-the-art studies (e.g., Bouwens et al.\ 2015).  The results
are shown in Figure~\ref{fig:lf8}, and it is clear that our result
lies somewhere midway between the $z\sim7$ and $z\sim8$ LFs, as one
might expect given the redshift distribution of the sources that make
up our $z=7.1$-8.5 sample.

\section{Summary}

In this paper, we take advantage of the deep Spitzer/IRAC observations
available over all five CANDELS in conjunction with the
HST+ground-based data to conduct a search over 900 arcmin$^2$ to find
bright $z\sim8$ galaxies.  To identify galaxies at such high
redshifts, we select those galaxies with especially red Spitzer/IRAC
[3.6]$-$[4.5] colors (i.e., $>$0.5), in the hopes of identifying those
$z\gtrsim6$ galaxies which show the presence of a strong
[OIII]+H$\beta$ line in 4.5$\mu$m band.  Such a selection is useful
for the CANDELS program, given the lack of uniformly deep $Y$-band
observations over all five fields.

Our selection yielded 4 $z\geq7$ candidates brighter than an
$H_{160,AB}$ magnitude of 25.5.  Each of these four selected
candidates was requried to be undetected ($<$2.5$\sigma$) at optical
wavelengths ($<$1$\mu$m), as defined by the inverse-variance-weighted
mean flux measurement, be undetected in the $V_{606}$-band
($<$1.5$\sigma$), and show a $I_{814}-J_{125}$ color redward of 1.5.

Fortuitously, 3 of our 4 selected $z\geq7$ candidates had deep
$Y$-band observations available from either deep ground-based
observations or from the new z9-CANDELS follow-up program (Bouwens
2014; Bouwens et al.\ 2016) with HST.  The available $Y$-band
observations provide clear confirmation of the $z\geq 7$ redshifts we
estimate for three of four candidates found in our search.

The redshift estimates we obtain for three of our selected candidates
lie significantly above $z\sim7$, with EGS-zs8-2 having a
redshift estimate of 7.6$\pm$0.3, EGS-zs8-1 having a redshift
estimate of 7.9$\pm$0.4, and EGSY-2008532660 having a redshift
estimate of 8.6$_{-0.4}^{+0.2}$.

We also obtained spectroscopic observations on two of our candidate
$z>7$ galaxies in the near-IR and find probable Ly$\alpha$ lines in
their spectra consistent with redshifts of 7.4770$\pm$0.0008 and
7.7302$\pm$0.0006.  The detection of Ly$\alpha$ emission for these
candidates is significant at 4.7$\sigma$ and 6.1$\sigma$ significance,
respectively.  The second of these sources was featured in Oesch et
al.\ (2015b).  Remarkably enough, a third candidate from our list was
spectroscopically confirmed to lie at $z=8.683$ by Zitrin et
al.\ (2015).

These sources represent the brightest $z\geq7.5$ candidates we
identified over the entire CANDELS program and are 0.5-mag brighter
than $z\geq7.5$ candidates identified anywhere else on the sky.
Coindentally enough, they all lie in the same CANDELS field, again
suggesting large field-to-field variations for brightest $z\geq7$
galaxies.  See also discussion in Bouwens et al.\ (2015) and Bowler et
al.\ (2015).

Using these candidates, we estimate the volume density of bright
($H_{160,AB}<25.5$) $z\geq7$ galaxies in the early Universe based on
our selected sample and estimate that 17$_{-11}^{+16}$\% of $z>7$
galaxies do not show such red colors.  The volume density estimate we
derive lies midway between the volume density of luminous $z\sim6$
galaxies Bouwens et al.\ (2015) derive and the volume density of
luminous $z\sim8$ galaxies.

The median [3.6]$-$[4.5] color distribution for our selection and
other bright $z\gtrsim7.5$ galaxies from the literature is
0.82$_{-0.20}^{+0.08}$ mag (observed) and 0.58$_{-0.20}^{+0.08}$
(correcting for the approximate selection bias: see \S6.1).  This
strongly points to the existence of extremely high EW nebular emission
lines in typical star-forming galaxies at $z>7$.  Assuming no
contribution from nebular line emission to the [3.6] band implies a
[OIII]+H$\beta$ EW of $\sim$800\AA.  However, allowing for
contamination of the [3.6] band in accordance with the expectations of
Anders \& Fritze-v.~Alvensleben (2003) would imply a median EW of
$\sim$1500 \AA.  These results are in reasonable agreement but perhaps
slightly higher than Smit et al. (2015) estimate for the IRAC-blue
sources they selected at $z=6.6$-6.9.  Smit et al.\ (2015) estimate a
typical $[OIII]+H_{\beta}$ EW of 1085\AA$\,$ for their selected
sources.  These estimates are similar albeit slightly higher than
those estimated by Labb{\'e} et al.\ (2013), Laporte et al.\ (2014,
2015), and Huang et al.\ (2016).

In the near future, we would expect the brightest $z\sim8$ galaxies to
be identified within the $\sim$1 deg$^2$ wide-area UltraVISTA field
(McCracken et al.\ 2012) by combining the progressively deeper
$YJHK_s$ observations with constraints from the optical Subaru+CFHT
observations and Spitzer/IRAC observations from SPLASH (Capak et
al.\ 2013) and SMUVS (Caputi et al.\ 2014).  Another significant
source of bright $z\sim8$ candidates will be the new BoRG$_{[z910]}$
program (Trenti 2014), which uses a huge allotment of 500 orbits to
cover a 500 arcmin$^2$ area to $\gtrsim26.5$ mag depth ($5\sigma$).

\acknowledgements

We thank Robert Barone-Nugent, Daniel Schaerer and Dan Stark for
valuable conversations.  This work has benefited significantly from
the public reductions of the SEDS program and hence the efforts of
Matt Ashby, Giovanni Fazio, Steve Willner, and Jiasheng Huang.  We are
grateful to Dan Stark, Sirio Belli, and Richard Ellis for
communicating with us with some unpublished results they also obtained
on EGS-zs8-2 (April 2015) where they also find a $>$3$\sigma$ line
(putatively Ly$\alpha$) at 1.031$\mu$m (to appear in D. Stark et
al. 2016, in prep).  We acknowledge the support of NASA grant
NAG5-7697, NASA grant HST-GO-11563, and a NWO vrij competitie grant
600.065.140.11N211.

\newpage
\appendix

\section{A.  Other Candidate $z\geq7$ Galaxies}

In addition to our application of our criteria to the catalogs Bouwens
et al.\ (2015) compiled over a 750 arcmin$^2$ search area within
CANDELS, we also made use of the catalogs from the 3D-HST team
(Skelton et al.\ 2014) over the same region.  Our rationale to do so
was to maximize the completeness of our selection for bright $z\geq7$
galaxies.

\begin{figure*}
\plotone{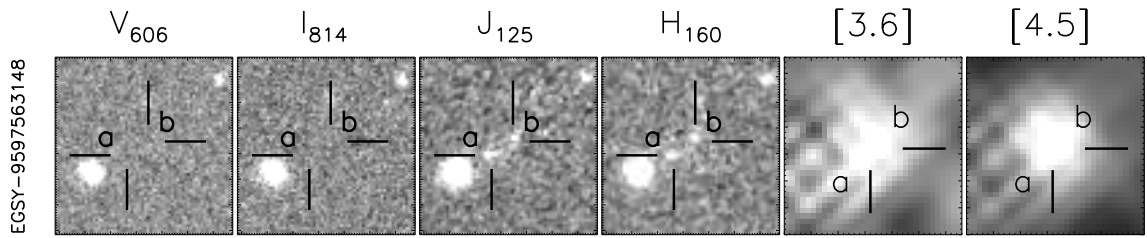}
\caption{HST/ACS $V_{606}I_{814}$ and HST/WFC3 $J_{125}H_{160}$
  postage stamp cut-outs ($4''\times4''$) of one particularly bright
  $(H_{160,AB}<25.5$) $z\geq7$ galaxy candidate selected by applying
  our IRAC-red criteria ($[3.6]-[4.5] > 0.5$) to the Skelton et
  al.\ (2014) photometric catalog.  This source also has a red
  $[3.6]-[4.5]$ color in the Bouwens et al.\ (2015) catalog, but do
  not quite satisfy our IRAC-red selection criterion.  This galaxy
  appears to consist of two separate components (indicated with black
  hash marks and the labels ``a'' and ``b'' respectively).  The
  photometry for each component is provided in
  Table~\ref{tab:table_details2}.\label{fig:postage_extra}}
\end{figure*}

\begin{figure*}
\plotone{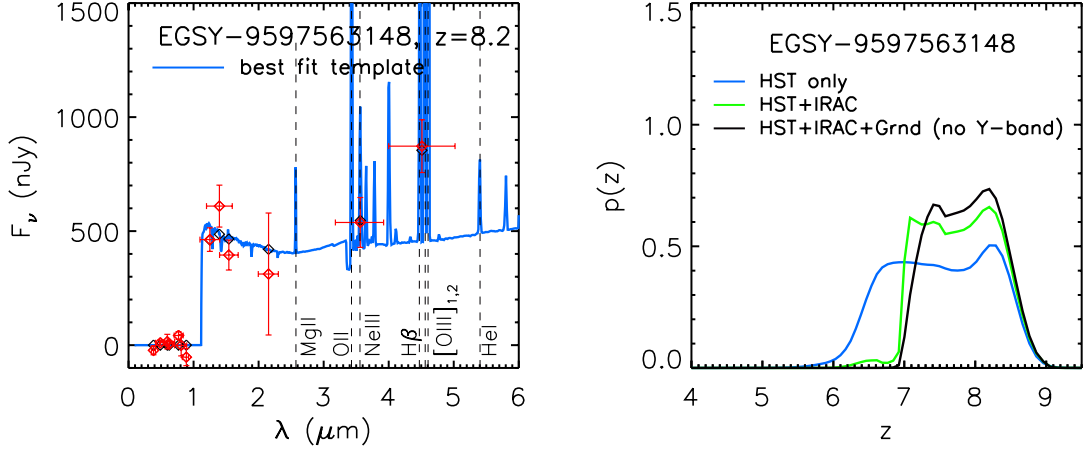}
\caption{\textit{Left}: Best-fit SED models (\textit{blue line}) to
  the observed HST + Spitzer/IRAC + ground-based photometry
  (\textit{red points and error bars}) for one especially bright
  $(H_{160,AB}<25.5$) $z\geq7$ galaxies identified by applying our
  IRAC-red selection criteria ($[3.6]-[4.5]>0.5$) to the Skelton et
  al.\ (2014) photometric samples.  \textit{Right}: The redshift
  likelihood distribution $P(z)$ for the same candidate $z\geq7$
  galaxy, as computed by EAZY using the Skelton et al.\ (2014)
  photometry.\label{fig:sed_extra}}
\end{figure*}

One additional $z\geq7$ galaxy candidate is found which did not make
it into our fiducial selection using the Bouwens et al.\ (2015)
catalogs (because it had a measured $[3.6]-[4.5]$ color of $\sim$0.2
mag).\footnote{While such differences might seem to be a concern, the
  [3.6]$-$[4.5] colors we measure for the 4 other sources in our
  selection agree to $<$0.1 mag with the Skelton et al.\ (2014) values
  ($\sim$0.05 mag differences are typical).}  We tabulate its
coordinates, $H_{160,AB}$-band magnitude, $[3.6]-[4.5]$ color, and
estimated redshift in Table~\ref{tab:table_details2}.  Postage images
of the $z\sim8$ candidate is presented in
Figure~\ref{fig:postage_extra}.  Model fits to the photometry Skelton
et al.\ (2014) provide on the source, as well as the inferred redshift
likelihood distribution, are also presented in
Figure~\ref{fig:sed_extra}.

\begin{deluxetable*}{ccccccc}
\tablewidth{0pt}
\tablecolumns{7}
\tabletypesize{\footnotesize}
\tablecaption{Additional bright $(H_{160,AB}<25.5$) z$\geq$7 source
  identified over the CANDELS fields utilizing the Skelton et
  al.\ (2014) catalogs for source selection.\label{tab:table_details2}}
\tablehead{
\colhead{ID} & 
\colhead{R.A.} & 
\colhead{Dec} &
\colhead{$m_{AB}$\tablenotemark{a}} & 
\colhead{$[3.6]-[4.5]$} & 
\colhead{$z_{phot}$\tablenotemark{b}} & \colhead{References\tablenotemark{*}}}
\startdata
EGSY-9597563148 & 14:19:59.76 & 52:56:31.40 & 25.03$\pm$0.10 & 0.53$\pm$0.26\tablenotemark{c} & 8.19$_{-0.87}^{+0.23}$ & [1]\\
$~~~$component-a & 14:19:59.78 & 52:36:31.30 & 25.73$\pm$0.14 \\
$~~~$component-b & 14:19:59.73 & 52:36:31.70 & 25.83$\pm$0.13 
\enddata
\tablenotetext{*}{References: [1] Bouwens et al.\ 2015}
\tablenotetext{a}{The apparent magnitude of each source in the $H_{160}$ band.}
\tablenotetext{b}{Maximum likelihood photometric redshift estimate from EAZY.}
\tablenotetext{c}{This is the [3.6]-[4.5] color that Skelton et al. (2014) measure.  Since our selection use the Spitzer/IRAC photometry from Bouwens et al.\ (2015) when available and Bouwens et al.\ (2015) measure a [3.6]-[4.5] color of $\sim$0.2 for the source, we do not include this source in our primary selection.}
\end{deluxetable*}

\section{B.  Sources Used to Validate our Proposed $[3.6]-[4.5]>0.5$ Selection}

In \S3.2, we considered a selection of sources from the four CANDELS
fields with deep $Y$-band observations to test the idea that we could
use an IRAC color criterion, i.e., $[3.6]-[4.5]>0.5$, combined with an
optical dropout criterion to identify galaxies at $z>7$ even in the
absence of $Y$-band data.

In Table~\ref{tab:valid_samp}, we provide a compilation of the 15
sources that we identified which satisfied the primary selection
criteria fom the paper but which are brighter than 26.7 mag in the
$H_{160}$-band (and brighter than 26.5 mag over the CANDELS UDS and
COSMOS fields).

\begin{deluxetable}{cccccc}
\tablewidth{10cm}
\tablecolumns{6}
\tabletypesize{\footnotesize}
\tablecaption{Sources in the 4 CANDELS fields with deep $Y$-band data used to validate our proposed selection technique (\S3.2)\label{tab:valid_samp}}
\tablehead{
\colhead{ID\tablenotemark{$\dagger$}} & 
\colhead{R.A.} & 
\colhead{Dec} &
\colhead{$m_{160,AB}$} &
\colhead{$[3.6]-[4.5]$} &
\colhead{$z_{phot}$$\ddagger$}}
\startdata
GNDY-6487514332 & 12:36:48.752 & 62:14:33.29 & 26.4 & $0.6_{-0.6}^{+0.7}$ & 7.66\\
GNDY-7048017191 & 12:37:04.805 & 62:17:19.14 & 26.2 & $1.1_{-0.2}^{+0.3}$ & 7.84\\
GNWY-7379420231 & 12:37:37.941 & 62:20:23.14 & 26.5 & $0.5_{-0.5}^{+1.5}$ & 8.29\\
GNWZ-7455218088 & 12:37:45.529 & 62:18:08.87 & 26.5 & $0.7_{-0.2}^{+0.2}$ & 7.16\\
GSDZ-2468850074 & 03:32:46.889 & $-$27:50:07.45 & 26.0 & $1.1_{-0.1}^{+0.1}$ & 7.24\\
GSWY-2249353259 & 03:32:24.934 & $-$27:53:25.94 & 26.1 & $0.6_{-0.3}^{+0.2}$ & 8.11\\
GSDY-2209651370 & 03:32:20.964 & $-$27:51:37.02 & 26.3 & $1.0_{-0.8}^{+1.8}$ & 7.84\\
COSY-0439027359 & 10:00:43.90 & 2:27:35.9 & 26.6 & $0.7_{-0.2}^{+0.2}$ & 7.33\\
COSZ-0237620370\tablenotemark{*} & 10:00:23.76 & 2:20:37.0 & 25.1 & $1.0_{-0.1}^{+0.2}$ & 7.14\\
COSY-0235624462 & 10:00:23.56 & 2:24:46.2 & 25.7 & $0.9_{-0.1}^{+0.1}$ & 7.84\\
UDSY-4133353345 & 02:17:41.333 & $-$5:15:33.45 & 25.8 & $0.5_{-0.2}^{+0.2}$ & 7.41\\
UDSY-4308785165 & 02:17:43.087 & $-$5:08:51.65 & 26.3 & $0.7_{-0.5}^{+0.8}$ & 7.84\\
UDSZ-4199355469 & 02:17:41.993 & $-$5:15:54.69 & 26.5 & $1.8_{-0.6}^{+2.6}$ & 7.08\\
UDSY-1765825082 & 02:17:17.658 & $-$5:12:50.82 & 26.3 & $1.1_{-0.1}^{+0.1}$ & 7.93\\
UDSY-5428621201 & 02:16:54.286 & $-$5:12:12.01 & 26.1 & $1.0_{-0.7}^{+1.1}$ & 7.49\\
\enddata
\tablenotetext{*}{Also featured in our primary selection.}
\tablenotetext{$\dagger$}{IDs are from the Bouwens et al.\ (2015) catalog.}
\tablenotetext{$\ddagger$}{Photometric redshift computed based on the HST and deep ground-based data (i.e., not including constraints from the IRAC data).}
\end{deluxetable}

\end{document}